\documentclass[manuscript,nonacm]{acmart}

\usepackage{xcolor}
\usepackage{ifthen}
\usepackage{appendix}
\usepackage{subcaption}
\usepackage{enumitem}
\usepackage{multirow}

\usepackage[most]{tcolorbox}

\usepackage{booktabs}
\usepackage{multirow}
\usepackage{xltabular}
\usepackage{longtable}
\usepackage[table]{xcolor}
\usepackage{tikz}
\usepackage{pifont}
\usepackage{makecell}
\usepackage{xspace}
\definecolor{ppgreen}{HTML}{0B6B3A}
\definecolor{pporange}{HTML}{B26A00}
\definecolor{ppgray}{HTML}{666666}

\newcommand{\htag}[1]{\textsuperscript{\textcolor{darkgray}{\texttt{#1}}}}

\newcommand{\ticket}[2]{%
    \mbox{%
        \tikz[baseline=-0.6ex]{
            \node[
                fill=black!45,
                text=white,
                rounded corners=2pt,
                inner xsep=3pt,
                inner ysep=1.2pt,
                font=\small\bfseries
            ] (id) { \texttt{#1} };
        }%
        \,#2%
    }%
}

\newcommand{\button}[1]{
    \tikz[baseline=(X.base)]{ 
        \node[ draw, rounded corners=1.5pt, inner xsep=3pt, inner ysep=1pt, font=\small\sffamily ] (X) {#1}; }%
        }

\usepackage{booktabs}
\usepackage{tabularx}
\usepackage{array}
\usepackage[table]{xcolor}

\definecolor{siggreen}{RGB}{235,247,237}

\definecolor{bigrow}{RGB}{242,242,242}
\definecolor{subrow}{RGB}{248,248,248}

\definecolor{hyplight}{HTML}{F2F3F5}
\definecolor{hypidbg}{HTML}{FFFFFF}

\newsavebox{\hypbgbox}
\newlength{\hypbgextra}
\newlength{\hypidwidth}
\usepackage{setspace} 

\newcommand{\summary}[1]{%
  \leavevmode\par
  \begin{tcolorbox}[
    width=\linewidth,
    colback=gray!4!white,
    colframe=gray!95!white,
    boxrule=0.5pt,
    arc=0pt,
    boxsep=0pt,
    left=7pt,
    right=7pt,
    top=6pt,
    bottom=6pt,
    before skip=5pt,
    after skip=5pt,
  ]
    {\setstretch{1.2}%
    \noindent\textbf{Summary:} #1\par}
  \end{tcolorbox}%
}

\newcommand{\hyp}[3][]{%
  \raggedright
  \hspace{-12pt}%
  \if\relax\detokenize{#2}\relax
    \colorbox{hypidbg}{%
      \begin{tikzpicture}[baseline=-0.5ex]
        \path[use as bounding box]
          (0em,-0.25em) rectangle (1.25em,0.25em);

        \begin{scope}[overlay]
          \if\relax\detokenize{#1}\relax
            \draw[line width=0.45pt, gray!60]
              (0.45em,0.85em) -- (0.45em,-0.85em);
          \else
            \draw[line width=0.45pt, gray!60]
              (0.45em,0.85em) -- (0.45em,-0.25em);
          \fi
          \draw[line width=0.45pt, gray!60]
            (0.45em,-0.25em) -- (1.15em,-0.25em);
        \end{scope}
      \end{tikzpicture}%
    }%
    \hspace{2px}%
    {\footnotesize #3}%
  \else
    \colorbox{hypidbg}{%
      \rule{0pt}{0.2em}%
      \textbf{\scriptsize\texttt{#2}}%
    }%
    \hspace{0.25em}%
    {\footnotesize #3}%
  \fi
}
\newcommand{\posbar}[2]{%
  \begin{tikzpicture}[baseline=-0.6ex,x=1cm,y=1cm]
    \draw[gray!35,line width=0.4pt] (-0.42,0) -- (0.42,0);
    \draw[gray!55,line width=0.4pt] (0,-0.045) -- (0,0.045);
    \fill[ppgreen!70] (0,-0.035) rectangle (#1,0.035);
  \end{tikzpicture}\,#2%
}

\newcommand{\negbar}[2]{%
  \begin{tikzpicture}[baseline=-0.6ex,x=1cm,y=1cm]
    \draw[gray!35,line width=0.4pt] (-0.42,0) -- (0.42,0);
    \draw[gray!55,line width=0.4pt] (0,-0.045) -- (0,0.045);
    \fill[ppgreen!70] (-#1,-0.035) rectangle (0,0.035);
  \end{tikzpicture}\,#2%
}

\newcommand{\nobar}[1]{%
  \begin{tikzpicture}[baseline=-0.6ex,x=1cm,y=1cm]
    \draw[gray!25,line width=0.4pt] (-0.42,0) -- (0.42,0);
    \draw[gray!45,line width=0.4pt] (0,-0.045) -- (0,0.045);
  \end{tikzpicture}\,#1%
}

\newboolean{showcomments}
\setboolean{showcomments}{true}   

\newcommand{\commentauthor}[3]{%
  \ifthenelse{\boolean{showcomments}}{%
    \textcolor{#1}{\textbf{#2's Comments:} #3}%
  }{}
}

\newcommand{\pp}{ParallelPilot\xspace}

\AtBeginDocument{%
  \providecommand\BibTeX{{%
    \normalfont B\kern-0.5em{\scshape i\kern-0.25em b}\kern-0.8em\TeX}}}
\setcopyright{cc}
\copyrightyear{2027}
\acmYear{2027}
\acmDOI{}
\acmISBN{}

\begin{document}

\title{ParallelPilot: Supporting Coordination and Monitoring in Parallel AI Coding}



\settopmatter{printacmref=false}
\renewcommand\footnotetextcopyrightpermission[1]{} 

\thanks{Correspondence to Tao Long <long@cs.columbia.edu>, Hussein Mozannar <hmozannar@microsoft.com>, and Weili Shi <weilishi@microsoft.com>}

\author{Tao Long}
\email{long@cs.columbia.edu}
\affiliation{
  \institution{Microsoft Research AI Frontiers, Columbia University}
  \city{New York}
  \state{New York}
  \country{USA}
}


\author{Weili Shi}
\email{weilishi@microsoft.com}
\affiliation{
  \institution{Microsoft Research AI Frontiers}
  \city{New York}
  \state{New York}
  \country{USA}
}

\author{Hussein Mozannar}
\email{hmozannar@microsoft.com}
\affiliation{
  \institution{Microsoft Research AI Frontiers}
  \city{New York}
  \state{New York}
  \country{USA}
}

\author{Maya Murad}
\email{mayamurad@microsoft.com}
\affiliation{
  \institution{Microsoft Research AI Frontiers}
  \city{New York}
  \state{New York}
  \country{USA}
}

\author{Rafah Hosn}
\email{rafahhosn@microsoft.com}
\affiliation{
  \institution{Microsoft Research AI Frontiers}
  \city{New York}
  \state{New York}
  \country{USA}
}

\renewcommand{\shortauthors}{Long et al.}

\begin{abstract}


As coding assistants become increasingly autonomous, developers run multiple sessions in parallel, shifting the challenge from code generation alone to coordinating and monitoring concurrent agent work. Through a formative study ($N=14$), we identified PILOT: five supervisory practices for Planning, Isolating, Logging, Observing, and Triaging parallel sessions. We present ParallelPilot, a design probe that instantiates PILOT through a planning interface, a run-logger, and an ambient dashboard alongside existing coding tools. In a counterbalanced within-subjects study ($N=16$), participants using ParallelPilot increased ticket throughput by 63\% in short coding tasks and supervised an average of one more concurrent agent at peak, while their tracking effort and context switching dropped. ParallelPilot also clarified execution plans, task dependencies, and intervention cues, and 14 of 16 participants preferred it over their current setup. These gains were not accompanied by significant improvements in perceived control or perceived success in redirecting the agents. Our findings demonstrate the value of explicit supervision support and position PILOT as a scaffold for designing tools that help people supervise concurrent work within and beyond coding. We suggest that future coding assistants should pair high-level awareness with low-cost paths back to the implementation evidence developers need to judge and steer agent work.

\end{abstract}



\begin{teaserfigure}
 \vspace{2px}
  \includegraphics[width=1.01\linewidth]{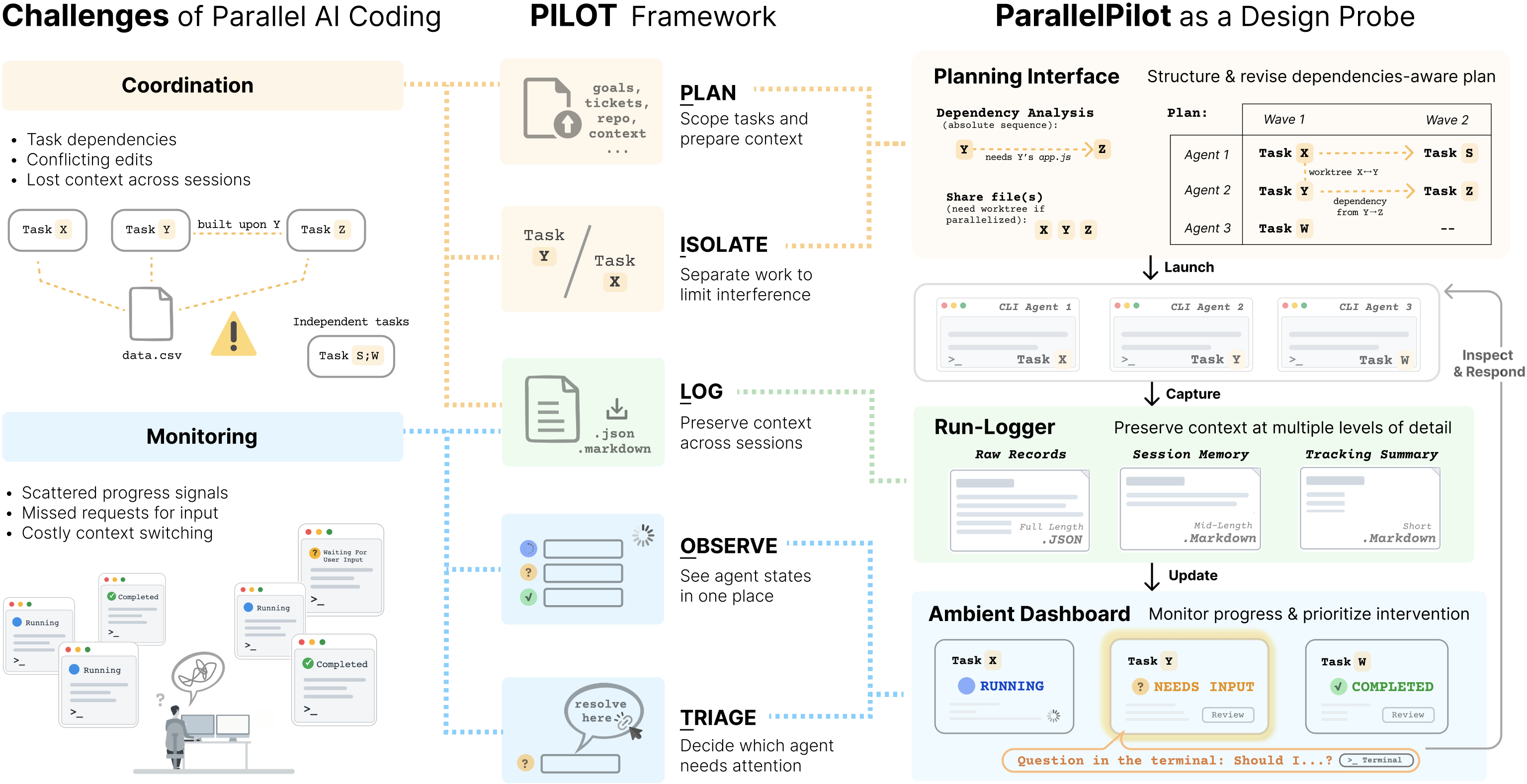}
  \caption{Overview of two challenges in parallel AI coding---coordination and monitoring (left)---the PILOT framework (center), and the ParallelPilot design probe (right). PILOT links five supervisory practices (Plan, Isolate, Log, Observe, Triage) to interface support structuring parallel work, preserving session context, and directing attention to agents that need input. ParallelPilot implements these practices through a planning interface, a run-logger, and an ambient dashboard, alongside existing coding tools.}
  \vspace{2px}
  \Description{
Conceptual overview diagram divided into three vertical sections: Challenges of Parallel AI Coding, PILOT Framework, and ParallelPilot as a Design Probe.
Left section: Challenges are split into Coordination (task dependencies, overlapping edits) and Monitoring (scattered progress signals, missed input requests, costly context switching). Example task dependency graph shows tasks X, Y, Z, S, W with shared files and conflict warnings.
Middle section: The PILOT framework, an acronym for Plan, Isolate, Log, Observe, Triage. PLAN structures tasks and context; ISOLATE identifies and separates dependent work; LOG preserves session context; OBSERVE monitors distributed progress; TRIAGE surfaces intervention cues. Dashed lines connect challenge problems to corresponding framework components.
Right section: ParallelPilot design probe implementation. A Planning Interface performs dependency analysis and defines wave‑based agent task plans. Agents launch to execute tasks. Run‑Logger preserves context across raw JSON, mid‑length Markdown, and short summary formats. An Ambient Dashboard visualizes agent status (running, needs input, completed). A closed‑loop human feedback path: Inspect \& Respond feeds back to the Planning Interface for iterative revision.
Color coding: beige for planning/coordination, light blue for monitoring/intervention. Dashed connectors link problem, framework, and prototype components.
}
\label{fig:pilot_overview}
\label{fig:teaser}
\end{teaserfigure}


\makeatletter
\let\@authorsaddresses\@empty
\makeatother

\maketitle
\section{Introduction}


Software development is undergoing a fundamental shift, not only in what code gets written or how fast, but in how developers work~\cite{oppotunty_se_changinglandscape}. 
AI coding assistants are evolving from autocomplete tools into autonomous agents capable of executing complete engineering tasks, including writing code, running tests, debugging failures, and proposing architectural changes~\cite{motivation_actionupcodingcapability,autonomous_sweagent}. As these agents become more capable, developers are increasingly experimenting with concurrent workflows, running multiple agent sessions on different tasks within and across projects~\cite{concurrent_openaireport,parallel_liu,parallel_haijun}.

We refer to this emerging practice as \textit{parallel AI coding}: running multiple independent AI coding sessions concurrently, typically as separate agent instances or windows. Developers use this approach to make progress on multiple independently actionable tasks or projects at the same time, rather than working through them sequentially or repeatedly switching between them~\cite{parallel_addcoordination14imporve, parallel_opportunty, parallel_cursor}. For example, a full-stack developer might set up three separate Claude Code sessions through the command-line interface (CLI) to handle three tickets: a frontend feature, a backend bug, and test writing. The developer monitors and coordinates all three concurrently, responding when the backend bug session asks for clarification while the others continue working. When the frontend session finishes, they review its result and send a follow-up to add a dark theme feature.

Parallel AI coding does not simply add more agents to the development process—it adds a new layer of coordination around them, creating substantial overhead~\cite{challenges_concurrent_dependencies,parallel_opportunty}. Developers must maintain a mental model of the parallel effort and respond to the events produced by each session. They need to decide how to divide work across sessions, what context each should receive, and which tasks must be completed before, after, or independently of one another. As work unfolds, they must also keep track of which session or terminal window corresponds to which task, how each task is progressing, and where intervention is needed.  This becomes increasingly difficult as sessions in existing coding interfaces generate continuous streams of logs, code changes, errors, and input requests. Developers are thus left to piece together the state of their growing fleet across terminals, windows, and logs.

In this paper, we focus on two intertwined forms of supervisory work in parallel AI coding, adapting concepts from traditional supervisory control theory~\cite{theory_coordination_monitoring,theory_coordination_monitoring2}. 
\textit{Coordination} involves establishing and maintaining the structure of parallel work, including task decomposition, dependencies, file boundaries, and agent assignments. \textit{Monitoring} involves tracking live agent activity, detecting completions, errors, stalls, and input requests, and determining which sessions require intervention.
We study these demands through their observable costs, including context switching, monitoring effort, and the need to recover or reconstruct session state after interruptions.

Through a formative study with 14 developers who regularly use parallel AI coding, we found that they already experience this dual coordination and monitoring burden. 
Most of them spent considerable effort deciding how to coordinate work across sessions and isolate them to avoid conflicts. To monitor the sessions, they developed ad hoc workflows combining terminal windows, markdown logs, custom dashboards, multi-monitor layouts, and other tools to track distributed agent state and surface issues. Yet these workarounds remained brittle and exhausting: developers missed when agents stalled or required input, lost track of which session was working on which feature, and struggled to recover context after interruptions. Across these practices, we identified five recurring strategies—\underline{P}lanning, \underline{I}solating, \underline{L}ogging, \underline{O}bserving, and \underline{T}riaging—which we synthesize as the PILOT framework.

We introduce \pp, a supervision cockpit designed to reduce the overhead of parallel AI coding. Rather than replacing developers' existing agent CLIs, editors, or workspace arrangements, \pp adds a lightweight coordination and monitoring layer around them. Its three components correspond to complementary aspects of PILOT: (1) a planning interface that supports dependency-aware task structuring, (2) a run-logger that captures per-agent events and traces, and (3) an ambient dashboard that synthesizes this information into a glanceable overview, surfaces sessions requiring attention, and provides summaries for rapid context recovery.

We evaluated \pp in a counterbalanced within-subjects study with 16 developers, comparing matched coding tasks with and without the tool. \pp improved objective task performance: participants completed more tickets, reached full completion more often, and achieved higher throughput. Participants also reported improved perceived efficiency, better planning ability, situational awareness, and ease of context recovery while comfortably supervising more concurrent sessions. However, these gains did not extend to perceived control or participants’ perceived success in redirecting the agent through their interventions.
Participants could keep up with parallel agent activity more effectively without necessarily feeling more confident in verifying or intervening on agent outputs. This highlights an important boundary of supervisory interfaces for parallel AI coding: making parallel work easier to coordinate and monitor does not, by itself, make agent work easier to verify or to intervene on.

Overall, we make three contributions:
\begin{itemize}[left=12px]
\item \textbf{The PILOT Framework}—an empirically grounded framework for parallel AI coding supervision from a formative interview study ($N=14$), identifying five pillars: Planning, Isolating, Logging, Observing, and Triaging. Together, these strategies help mitigate the costs of coordination and monitoring.

\vspace{3px}

\item \textbf{ParallelPilot}—a supervision cockpit that operationalizes PILOT through three components: dependency-aware planning, run logging, and an ambient dashboard for glanceable monitoring and actionable triaging. 

\vspace{3px}

\item \textbf{Empirical insights from a controlled within-subjects study} ($N=16$) providing evidence of higher measured ticket throughput and lower self-reported tracking and context-switching burden in short parallel coding tasks, while perceived control and intervention success did not significantly change. Interview findings motivate a distinction between high-level progress awareness and the implementation context needed for steering.

\end{itemize}

\section{Related Work} 

\subsection{Managing Concurrent Coding Tasks and Coordination Challenges}

Managing concurrent streams of coding tasks promises greater efficiency in task completion~\cite{parallel_github}. Studies and reports show that a single developer can now achieve output that previously required multiple engineers: shipping 72 story points per day and 22 PRs per week~\cite{parallel_stoa}; using four AI agents to complete projects in half the time of a linear one-at-a-time development workflow~\cite{parallel_onedevAI_do4times}. Cursor reported that parallel AI coding enables doubling weekly PR throughput and compressing an 18-month migration into a single engineer running a fleet of agents~\cite{parallel_cursor}. However, managing concurrent tasks and AI tools imposes two major demands: effective coordination and continuous monitoring~\cite{parallel_opportunty}. We discuss the coordination challenge below and the monitoring challenge in Section~\ref{sec:2.2}. 


Concurrent work scales up the amount of information people must coordinate at once. Developers must maintain an understanding of each task's dependencies, priorities, conflicts, progress, timeline, needed context, relationships to the larger project, and past or future activities. 
Coordination then operates on two levels: selecting an initial set of tasks and iteratively adapting as newly available capacity emerges.
Thus, it requires decomposing and sequencing work into manageable, parallelizable components when tasks are set together.
In engineering, tools like GitHub and designs like Gantt charts help externalize this structure through issue trackers, branches, pull requests, project boards, kanban, continuous integration systems, and version control~\cite{parallel_github}. Previous agentic tools introduce structural orchestration and a separate orchestrator agent to assist with coordination~\cite{agent_orchestration,agent_orchestration2, agent_orchestration3}. A recent study~\cite{parallel_addcoordination14imporve} shows parallel AI coding tools with stronger dependency-finding and coordination design lead to a 14\% improvement in pass rate, a 2.10× wall-clock speedup, and a 35\% reduction in API cost. Distributed cognition~\cite{distributedcognition_new, distributedcognition_old} also provides coordination scaffolding. Hutchins' cockpit example~\cite{distributedcognition_old} illustrates how pilots manage and coordinate concurrent threads of information—altitude, speed, and communications—with instruments and ground control through shared states, displays, and adaptive procedures. DoubleAgents, a recent HCI tool powered by parallel AI agents~\cite{distributedcognition_doubleagents}, adapts this framework by offloading the event organizer's coordination efforts through parallel agents, visualizing work-thread progress, and codifying the organizer's coordination strategies as future alignment materials.
Beyond structure, Vasilescu et al.~\cite{parallel_github} identify context switching as the key cost of parallelization. The structural complexity and lack of traceability of concurrent threads create coordination barriers, as developers must mentally switch contexts and recall relevant background information to understand other threads when viewing them with fresh eyes~\cite{parallel_github,agent_comm_challenge}. Developers now create developer journals to document tasks so they can reference tangible trace artifacts and context-switch effectively among sessions~\cite{contextswitching_log,developer_log}. However, as these AI systems become more autonomous and proactive, structuring and preparing work for parallel coding may require additional support.




\subsection{The Monitoring Challenges of AI Coding Assistants: Legibility and Human Factors}
\label{sec:2.2}
AI coding assistants produce code, logs, and explanations at overwhelming volumes and speeds, creating a legibility problem for human monitoring~\cite{oversight_log_hard_to_read_Madeleine,loghardtoread,viz_hardtoreadfromlongconversationwhendebuggingmultiagent,viz_traceishardtounderstandCUA}. Anthropic reported that one single user task generates roughly 12 Claude actions and 3,200 words~\cite{claudeoutputMuch}. Generation is 5–7× faster than humans can read or comprehend~\cite{log_faster5xhumanreadability}. A recent quantitative study~\cite{log_tracelab_cantread} echoed that coding agents generate output at 33.9--46.8 tokens per second, compared with roughly 5.3 tokens per second\footnote{Converted from the estimated average silent reading speed of 238 words per minute for adults reading English non-fiction~\cite{readingspeed_238}, using the standard approximation of 1 token $\approx$ 0.75 English words~\cite{openai_understanding_nodate}.} for average adult reading speed~\cite{readingspeed_238}.
Beyond intermediate logs, final code generation is also overwhelmingly fast: AI agents average 140–200 lines of meaningful code per minute, whereas focused humans read 20–40 lines~\cite{log_faster5xhumanreadability}. 

Even when coding assistants make their processes readable, developers may still not notice, read, or understand what the agent is doing. A prior HCI study~\cite{log_develoeprNotRead} shows that developers frequently skim or disregard lengthy execution traces to avoid substantial cognitive demands and information overload. Other studies~\cite{overtrust_aicodingagents, overtrust_morecode_lessverification} show that such skimming is common among developers who overrely on coding assistants, perceiving them as sufficiently competent or as a convenience.
However, issues frequently arise when users miss critical intervention points due to insufficient monitoring~\cite{agent_comm_challenge,visibility_challeneg_facct}. An empirical study of over 100 developers collaborating with coding agents~\cite{overtrust_notreading_AISabotage} found that 94\% failed to detect agent-inserted sabotage; even when explicit monitor warnings were present, 56\% accepted malicious code, reporting that they ``\textit{didn't dwell on}'' the alerts. The identified failure modes are minimal code review (67\%), plausible cover stories (22\%), and overtrust built from routine use (11\%). Situational awareness theory~\cite{situationalawareness} emphasizes that the ability to detect and be aware of potential problem cases is a prerequisite for user control and intervention. The human-AI oversight community~\cite{oversight_workshop} also noted that users often struggle with ``\textit{recognizing critical situations or AI behavior}'' and ``\textit{executing timely interventions}''. 
Recent HCI efforts target this by building structural visualizations of agent traces~\cite{oversight_log_hard_to_read_Madeleine, visibility_example}, centralized tiered dashboards for monitoring multi-agent execution~\cite{viz_hardtoreadfromlongconversationwhendebuggingmultiagent, visibility_example2}, and proactive systems that decide when an agent should ask for user attention~\cite{proactivity_valerie, proactivity_mozannar2024show}.
Many developers constrain the intermediate logs and unnecessary explanation in output using \texttt{\small AGENTS.md} file or instructions to save tokens and improve readability~\cite{log_developeraddAgentMd}.  

Running multiple parallel AI coding assistants compounds monitoring challenges—e.g., enabling user interaction without disrupting ongoing processes, and allowing prioritization or interruption across agents~\cite{parallel_opportunty}. A six-week deployment of three concurrent research agents confirms that parallelizability amplifies the oversight burden~\cite{loghardtoreadoversight_produceoverwhelminginformation_paralleliz_ability_amplifies_the_oversight_burden}; the authors cite ``overwhelming information'' and users' inability to manage it, prompting the categorization of system traces to aid sense-making. Three week-long deployments of DoubleAgents similarly found that coordination-agent monitoring support reduced review effort and saved time, even for the task experts~\cite{distributedcognition_doubleagents}. This motivates the need for support that helps developers track parallel agents, intervene selectively, and avoid overreliance on automation.

\section{Formative Study: Understanding How Developers Supervise Parallel AI Coding}

\subsection{Overview}

To characterize how experienced developers coordinate and monitor multiple concurrent AI coding sessions, we conducted a 60-minute formative interview study with 14 developers. 
We investigated their motivations for adopting parallel AI coding and their strategies for managing such a process.
The formative study was guided by two RQs:

\begin{itemize}
    \item RQ1: Why, when, and how often do developers turn to parallel AI coding?
    \item RQ2: What strategies and practices do developers use to coordinate and monitor parallel sessions?
\end{itemize}

\subsection{Participants}

We recruited 14 participants (P1--P14) through an internal mailing list for researchers and developers within our research organization. The recruitment message invited developers who regularly use AI coding assistants and have extensive experience running more than two AI coding sessions in parallel. The 14 participants (4 women, 10 men) spanned roles comprising 7 PhD research interns, 3 software engineers, 2 senior researchers, and 2 engineering managers. Seven participants were in the 25–34 age range, three in the 45–54, two in the 18–24, and one each in the 35–44 and 65+.

All participants were highly experienced coders: six participants reported more than 10 years, three reported 7–10 years, and five reported 4–6 years. All participants used AI coding assistants more than three times per day in their general workflow and demonstrated substantial seniority and expertise in parallel AI coding workflows---ten reported running multiple AI coding sessions in parallel more than 30 times, while four reported doing so 10–30 times.

\subsection{Procedure}
Each participant took part in a 60-minute semi-structured interview, either in-person or via Microsoft Teams. Before the interview, participants completed a short demographics and AI experience survey. 
During the interview, we asked participants to describe their current parallel AI coding workflows and invited them to screenshare and walk through a real project or setup in which they had used concurrent agents. We asked participants to map out the sessions running and what each one was working on. We probed how participants tracked each session: what they actively watched versus what they skimmed or ignored, what signaled that a session needed attention, how they judged a session's output, and how they knew a session was done. See Appendices~\ref{app:formative-survey-questions} and \ref{app:formative-interview-protocol} for the survey and interview questions.

All participants provided informed consent to participate and to have the interview recorded for transcription and analysis. They were compensated 50 USD, and the study protocol was approved by the internal ethics committee.

\subsection{Data Collection and Analysis}
Our dataset comprised the survey responses, interview recordings and transcripts, and screen-shared walkthroughs described above. We analyzed survey responses descriptively to characterize participants' experience and parallel coding practices. We used inductive affinity diagramming~\cite{affinityDiagramming,thematicAnalysis} across transcripts, notes,
and screen-shared observations to identify recurring patterns in
task decomposition, session organization, and monitoring. Codes
and themes were refined through multiple rounds of feedback and
discussion with all co-authors, informing a preliminary framework
for supervising parallel coding workflows.

\subsection{Findings}


\subsubsection{\textbf{RQ1: Why, when, and how often do developers turn to parallel AI coding?}}



\summary{\small Participants used parallel AI coding to fill workflow downtime, such as waiting for builds or tests. Participants reported parallel AI accounted for nearly as much coding as single-session AI (45.3\% vs.\ 48.6\%). CLI interfaces dominated, Copilot and Claude/Claude Code were the most common tools, and three concurrent sessions were typical.}

\begin{figure}[!h]
    \centering
    \vspace{-10px}
    \includegraphics[width=1.01\linewidth]{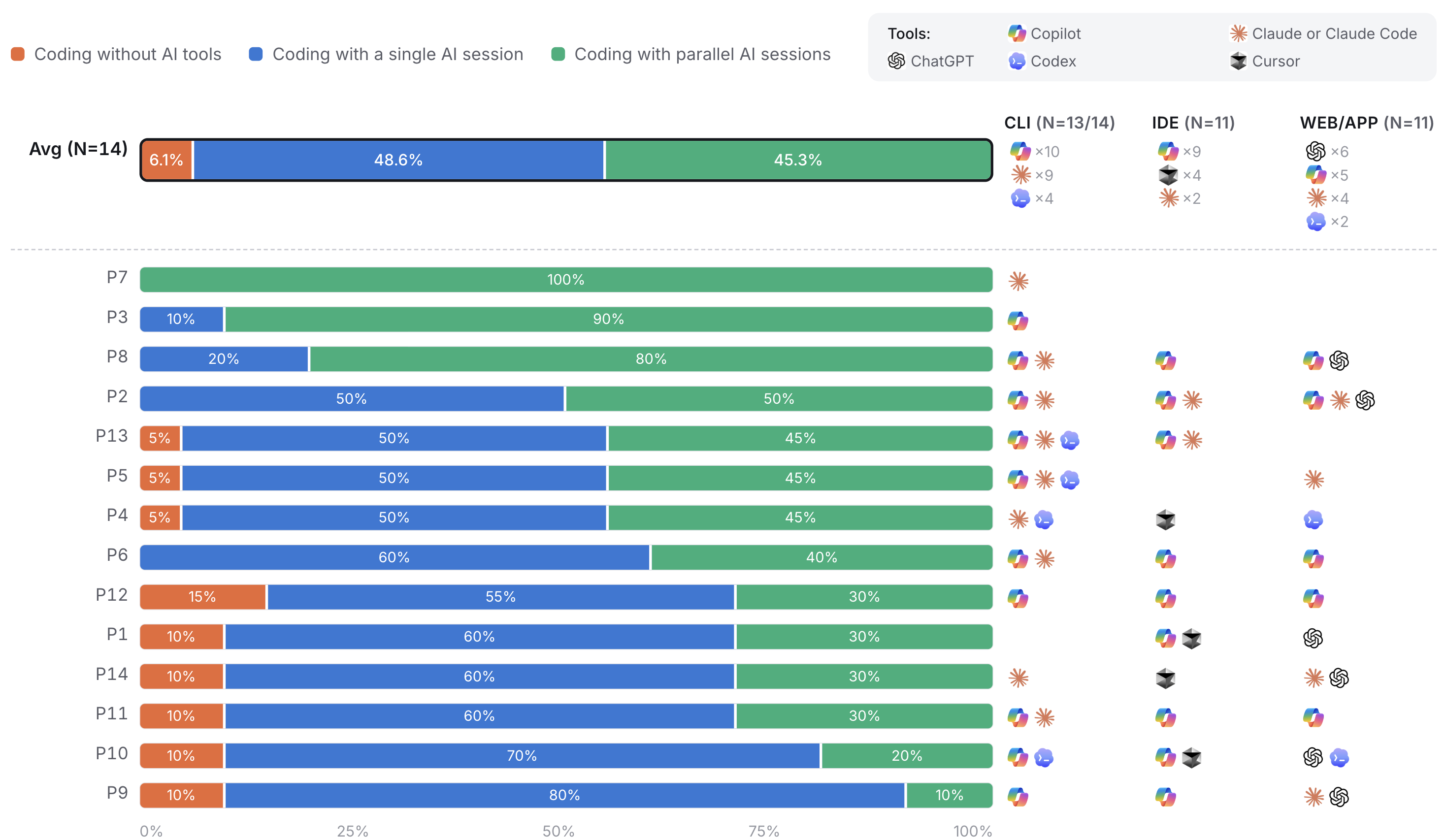}
    \caption{Current coding workflow distribution self-reported by participants ($N=14$). Each bar shows the percentage of coding time spent with no AI tools, a single AI session, or multiple AI sessions in parallel. Participants are sorted by parallel AI usage (highest to lowest); the group average is shown at the top, separated by a dashed line. Three additional columns on the right indicate which tools each participant uses across CLI, IDE-integrated chat interfaces, and WEB/APP (standalone web/desktop AI applications), with aggregate usage counts displayed in the average row.}
\Description{Horizontal stacked bar chart of 14 participants' coding workflow distribution across three categories: no AI tools (orange), a single AI session (blue), and parallel AI sessions (green). Bars are ordered from highest to lowest parallel AI use (P7 at 100\% down to P9 at 10\%). The group average, shown at the top, is 6.1\% no AI, 48.6\% single AI, and 45.3\% parallel AI. Three columns to the right show tool usage per participant across CLI, Web/App, and IDE interfaces.}
    \label{fig:workflow}
    \vspace{-20px}
\end{figure}

\paragraph{\textbf{Current Coding Workflow Distribution and Setup.}}
As shown in Figure~\ref{fig:workflow}, participants estimated how their current coding practice was distributed across three modes: 
AI-assisted coding was divided almost evenly between single-session AI ($M=48.6\%$) and parallel multi-session AI ($M=45.3\%$), with only 6.1\% of coding occurring without AI.

CLI-based interfaces were the dominant setup for AI-assisted coding, mentioned by 13 of 14 participants (92.9\%), followed by IDE-integrated chat interfaces and standalone web/desktop AI applications, each used by 11 participants (78.6\%). In terms of tools, Copilot was the most ubiquitous, appearing across 11 participants (78.6\%), followed by Claude or Claude Code (10 participants, 71.4\%). ChatGPT was reported by 6 (42.9\%), while Codex and Cursor were each reported by 4 (28.6\%). P8, P3, and P7 were heavy users of parallel AI workflows, allocating 80\%--100\% of their coding time to parallel AI, primarily using Claude Code and Copilot CLI.

Working with three AI coding sessions at once was the most common setup, reported by 8 of 14 participants. Most participants worked with two to four concurrent AI coding sessions. The maximum number of sessions participants had ever run at once ranged from three to eight; five participants reported five, and one reported eight. One participant, P6, described a cloud-based workflow that allowed them to delegate small, isolated issues across 20 sessions.

\begin{figure}[!b]
    \centering
    \begin{subfigure}[c]{0.65\linewidth}
        \centering
        \includegraphics[width=\linewidth]{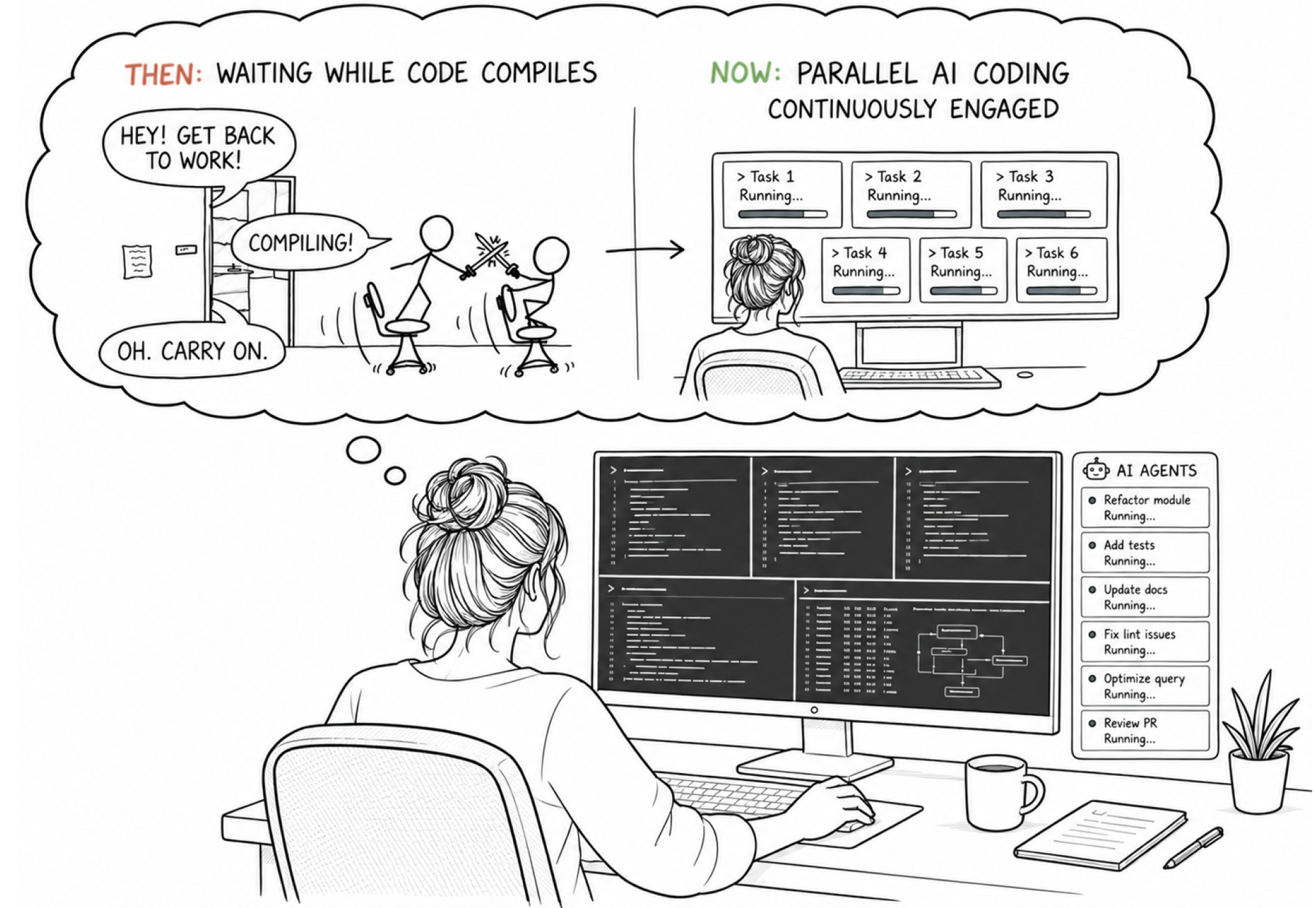}
        \caption{P5's reflection on parallel AI coding.}
        \label{fig:participant-comic}
    \end{subfigure}
    \hfill
     \begin{subfigure}[c]{0.31\linewidth}
        \centering
        \includegraphics[width=\linewidth]{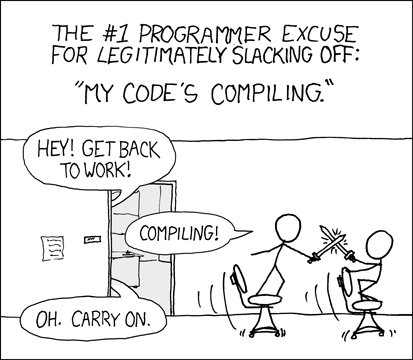}
        \caption{xkcd \#303, ``Compiling.''}
        \label{fig:xkcd-compiling}
    \end{subfigure}
    \vspace{-5px}
    \caption{P5 invoked the familiar developer trope depicted in xkcd \#303 to describe how parallel AI coding changes the experience of waiting. Instead of disengaging while code compiles, developers remain actively engaged by monitoring, answering, and redirecting multiple concurrent AI sessions. Source: Randall Munroe, xkcd \#303, \url{https://xkcd.com/303/}, licensed under CC BY-NC 2.5.}
    \Description{Two-panel illustration. Left: a participant shown from behind has a thought bubble containing a blank comic area and the text ``NO LONGER THIS... BACK TO FULL-TIME ENGINEERING,'' illustrating how parallel AI coding replaces idle waiting with continuous interaction with AI coding agents. Right: the xkcd ``Compiling'' comic depicts two programmers sword-fighting while their code compiles, representing legitimate idle time. }
    \label{fig:formative-comic}
    \vspace{-15px}
\end{figure}

\paragraph{\textbf{Downtime as a Trigger for Parallel AI Coding.}}
Beyond how often participants ran parallel sessions, our data point to a consistent trigger for \textit{when} they did so: idle time in their workflow---periods when they were waiting for a single agent to complete. P9 stated, ``\textit{I don't wanna wait sequentially for things to happen,}'' while P6 noted, ``\textit{[now] I can actually work on multiple things at once [...when] I have downtime or I'm waiting for things.}''

Participants described parallel AI coding as transforming the familiar developer experience of waiting for tools to finish. P5 shared that ``\textit{parallel AI coding is useful to do while you're waiting for something... having something else going is just nice.}'' As illustrated in Figure~\ref{fig:participant-comic}, P5 contrasted this with the classic developer trope of being idle while code compiles, referencing the xkcd comic ``Compiling'' (Figure~\ref{fig:xkcd-compiling}): ``\textit{It will no longer be like that [comic] with the still compiling... I can't be like, `Oh, I don't need to doomscroll. My code's working for me.}' ... \textit{Now when I'm using Claude Opus, running five or six things at a time, [I'm] pretty much always answering questions from the machine or trying to redirect... so it really becomes like you're doing full-time coding again.}'' (P5) 
This illustrates how parallel AI coding converts idle waiting time into an active orchestration loop. 
P3 described this as an explicit scaling strategy: ``\textit{I feel I want to scale by running multiple concurrent sessions... trying to keep them and myself busy}.''



\subsubsection{\textbf{RQ2: How Do Developers Coordinate and Monitor Parallel AI Coding?}}

\summary{Parallel AI coding required active supervision: all 14 participants reported coordination and monitoring overhead. We introduce their management strategies here as PILOT; Section 4 develops each strategy with evidence.}


Downtime made parallel AI coding possible, but participants emphasized that making it useful required active supervision. All 14 participants described a real coordination and monitoring tax. P2 put it plainly: ``\textit{after a while [running concurrent sessions]... if it's this feature and that's that feature, it'd be over your brain, fried quickly.}'' The tax showed up as a recall cost too: P6 reported that they ``\textit{couldn't even remember what I had told Copilot}'' across sessions. P12 described feeling ``\textit{like a quarterback across three sessions},'' relying on peripheral glances and occasionally missing alerts when absorbed in one thread, calling the work ``\textit{a little more tiring}.'' P7 found long logs ``\textit{overwhelming}'' and asked agents for high-level markdown summaries instead, and P4 limited themselves to a single screen just to ``\textit{keep track}.''

Participants adapted their workflows to manage these demands, revealing five recurring strategies: \textbf{\textit{\underline{P}lanning}} what should happen in parallel, \textbf{\textit{\underline{I}solating}} sessions to prevent interference, \textbf{\textit{\underline{L}ogging}} context to preserve continuity, \textbf{\textit{\underline{O}bserving}} agent state without constant inspection, and \textbf{\textit{\underline{T}riaging}} situations requiring human intervention. We synthesize these strategies into the \textbf{PILOT} framework. 
The following section continues our answer to RQ2 by grounding each strategy in participants' accounts and concrete workflow examples.

\section{The PILOT Framework}

\begin{figure}[!b]
\definecolor{planningbg}{HTML}{FFFAF2}
\definecolor{loggingbg}{HTML}{F2FCF4}
\definecolor{dashboardbg}{HTML}{EDF9FF}

\DeclareRobustCommand{\componentbox}[2]{%
  \begingroup
  \setlength{\fboxsep}{1pt}%
  \smash{\colorbox{#1}{#2}}%
  \endgroup
}
    \centering
        \vspace{-5px}
    \includegraphics[width=.95\linewidth]{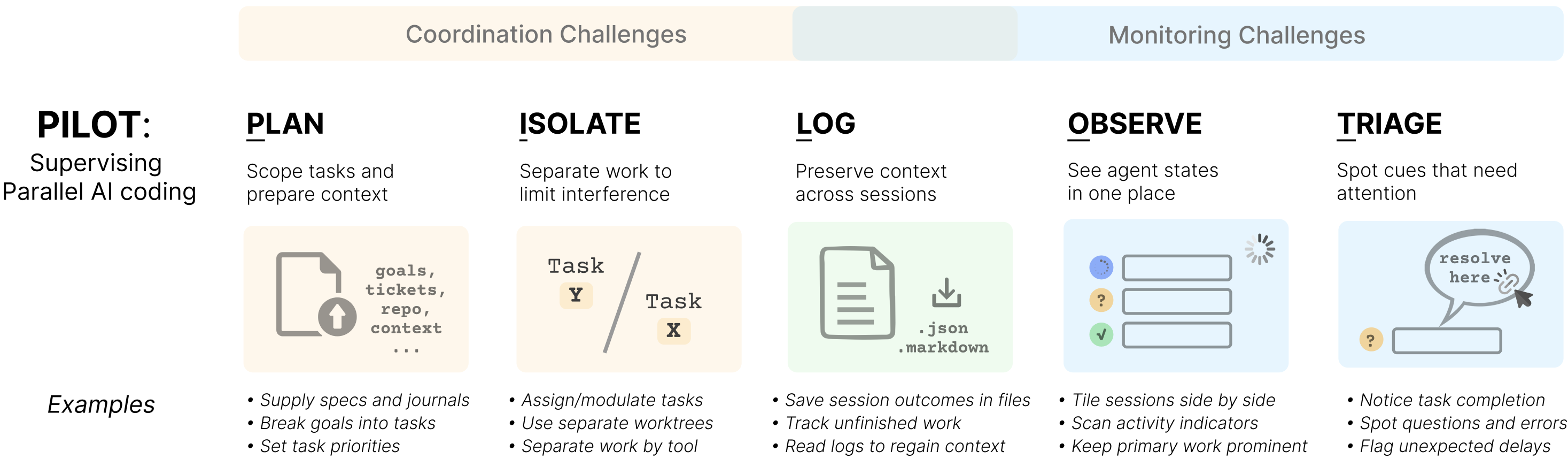}
    \caption{The PILOT framework for supervising parallel AI coding. Five complementary practices address \componentbox{planningbg}{coordination} and \componentbox{dashboardbg}{monitoring} challenges: \componentbox{planningbg}{Planning}, \componentbox{planningbg}{Isolating}, \componentbox{loggingbg}{Logging}, \componentbox{dashboardbg}{Observing}, and \componentbox{dashboardbg}{Triaging}. Logging bridges \componentbox{loggingbg}{both challenge areas} by preserving context across sessions. Examples beneath each pillar illustrate developer practices; the pillars are not a fixed sequence.}
\Description{A five-column diagram of the PILOT framework. Overlapping bands group Planning and Isolating under coordination challenges, Observing and Triaging under monitoring challenges, and Logging under both. Planning scopes tasks and prepares context through specs, task decomposition, and priorities. Isolating limits interference through separate files, worktrees, and tools. Logging preserves context by saving outcomes, tracking unfinished work, and rereading logs. Observing brings agent states into one view through tiled sessions, activity indicators, and a prominent primary task. Triaging helps developers notice completion, questions, errors, and unexpected delays. Each column includes an illustration and three example practices.}
    \label{fig:pilot}
    \vspace{-20px}
\end{figure}

As shown in Figure~\ref{fig:pilot}, the PILOT framework characterizes the strategies that help developers manage parallel AI coding: \textbf{\textit{\underline{P}lanning}}, \textbf{\textit{\underline{I}solating}}, \textbf{\textit{\underline{L}ogging}}, \textbf{\textit{\underline{O}bserving}}, and \textbf{\textit{\underline{T}riaging}}.
As the number of concurrent agents scales up, developers must not only \textit{coordinate} the work around them---deciding what can run in parallel, preventing sessions from interfering with one another, and preserving context across interactions---but also \textit{monitor} ongoing work, maintain awareness of agent progress, and decide when and where to intervene. 

We call the framework {PILOT} to evoke how airplane pilots coordinate and monitor through cockpit supports rather than track everything themselves, drawing on Hutchins’s account of distributed cognition~\cite{distributedcognition_old}.
As AI agents take on more of the direct programming work, developers increasingly pilot a portfolio of concurrent agent work threads. PILOT does not impose a fixed sequence: developers may move between and revisit earlier pillars as task and agent states change. Though derived from parallel AI coding, PILOT points to a broader problem of distributing limited human resources (time, attention, effort, etc.) across multiple concurrent activities by structuring and isolating work, externalizing context, maintaining situational awareness, and selectively allocating attention where needed.



\subsection{\textbf{P --- Planning: Scoping and Preparing the Work}}

Once downtime created an opening for parallel work, participants did not immediately launch multiple coding sessions. Instead, they scoped the work, established priorities, and prepared the agent context and scaffolding.

Participants first identified a manageable set of tasks to run concurrently by breaking larger, ambiguous goals into smaller units. 
Participants commonly identified a primary task for closer attention alongside secondary tasks that could be checked later, giving them a tentative agenda for where to look first as multiple sessions progressed. 
Then, participants prepared the context that agents needed before launch. This included specifying preferred approaches, relevant files and folders, constraints, and success criteria. P2 described a repeatable handoff containing ``\textit{spec, prior state, plan, and work done},'' alongside a journal preserving all past decisions. Several participants provided tickets, PR links, and background notes to help agents quickly orient themselves (P3, P4, P6, P12). Nine participants developed reusable artifacts to improve consistency across sessions: six described encoding preferences into pipelines and markdown files for uploading, while three others built them as skills. P13 explained, ``\textit{most of my skills are for repetitive tasks...}'' P2 used an issue-driven development \texttt{\small AGENTS.md}, while P11 added a ``grill me'' prompt to challenge and refine their plans.

Together, these practices show,
rather than increasing concurrency by simply opening more sessions, participants deliberately prepared the work and its surrounding context so that parallel execution could remain manageable.

\subsection{\textbf{I --- Isolating: Managing Separability}}

Once participants decided what to delegate, they had to keep parallel sessions from interfering with one another. P6 explained that parallelism became useful ``\textit{once a project has enough space for non-overlapping things.}'' 

To exercise separability, 11 of 14 participants modularized code and tasks, splitting work so that agents would not touch each other's outputs. 
P7 looked for ideas ``\textit{not running to the same file}'', P13 ran ``\textit{three very parallel threads, which did not rely on each other}'', and P9 split along subtasks: ``\textit{if it's a different subtask of the same project, I still have some kind of parallel workflow}.'' Many participants also relied on version control primitives. P3 used worktrees and branches so that concurrent sessions would not ``\textit{mess with other people's branches}''; P2 kept to one branch at a time, since ``\textit{multiple branches gets too confusing too quickly}''. Some participants even separated by tool use beyond code: P10 split work by using ``\textit{ChatGPT only for proofs, Codex to implement experiments}'', while P14 started a fresh conversation whenever pollution felt likely, ``\textit{I try less to pollute the whole conversation and make another session}.'' 
The other three sat at opposite ends. P8 and P4 cautiously isolated at the whole-project level, afraid they would ``\textit{pollute each other's context and make conflicting changes},'' whereas P11 skipped boundaries, accepting overlap and resolving merge conflicts later.


These practices show that successful parallelism depended on separability: tasks were independent enough that agents would not overwrite, duplicate, or conflict with each other. Misjudging it was costly: P2 found that overlapping work in the same repository cost ``\textit{more time later reconciling all of the differences}'' than doing the work sequentially.


\subsection{\textbf{L --- Logging: Externalizing Memory}}
Moving between multiple sessions made it difficult to remember what each agent had been asked to do, what it had tried, and why decisions were made, especially as AI histories and reasoning could quickly become inaccessible. P6 described this breakdown: ``\textit{I could [now] focus on the details of one [window] ... and then I couldn't remember or even find what I had told Copilot to do in the first window by the time I got back to it.}''

To manage this burden, participants borrowed from traditional developer journaling, externalizing their memory and AI session history to persistent documents on local computers or in the cloud.
Seven participants built markdown or text files, logging decisions and outcomes of the sessions, so they could check for themselves and later feed them back into the next session as context. P4 used files such as \texttt{\small NOTEBOOK.md} and \texttt{\small RESULTS.md}; P2 kept per-issue journals; P7 maintained one markdown file per task with plans, attempts, progress, and remaining work. P11 described reading these documents to recover state: ``\textit{I read quite a bit of markdown files... I'm not looking at code.}'' 

These logs were not passive records but active coordination devices: a spec could remind the developer of scope, support later verification, and provide context for steering future sessions. 
By offloading session context to durable artifacts, developers reduce the working memory cost of switching between sessions, recover context faster after interruptions, and promote a sense of security.


\subsection{\textbf{O --- Observing: Assembling a Centralized View}}
After launching parallel agents with logging in the backend, participants built  makeshift control rooms out of terminals and side panels across one or two extended monitors alongside their laptop screen  to observe raw agent states.

While screen setups varied, ten of the 14 participants arranged every active session so that a single glance could reach it—{an active-monitoring full screen}, requiring no active switch to check. Five of them split their main screen to keep two to four parallel sessions visible side by side, catching most status changes by scanning across them, though the windows shrank enough that fine detail was easy to miss. Two CLI-only users stacked terminal windows into one view on their main screen and watched the tab spinner in the navigation bar to determine their status. Three ran an IDE alongside a stacked chat or CLI panel in their main screen, keeping a full-detail primary session identified during \textit{planning} apart from peripheral ones they could absorb without switching windows. The remaining four gave each agent its own full screen, swapping screens entirely to check on an agent or closing one agent window to open another---a deliberate act that marked a task switch. Though built around each user's habits, they all shared the same cost: turning away from the current task, even briefly, often led to missed agent updates.

Across all participants, we see clear benefits of centralization: a single point of contact for every agent—one place to both check status and intervene when needed. Centralization lowers the manual cost of switching between screens and focuses user attention and potential steering more effectively.

\subsection{\textbf{T --- Triaging: Prioritizing Intervention Cues}}

Even with a centralized view, participants found it difficult to determine {which session needed attention now}. Important moments could still be buried in long logs and scattered across windows.
Triage therefore involved continuously {detecting which session had become actionable and deciding what required intervention}. In participants' current practice, triage often relied on lightweight, improvised observation. Participants turned their heads or clicked between sessions, watched for spinners or activity cues (e.g., the Claude logo or Copilot's blinking dot) to stop, and periodically discovered that an agent had been waiting for a question no one had answered. For longer-running or less observable tasks, they constructed additional signals, such as checking log statistics, generated files, dashboards, or verification outputs at intermediate points. Some even built lightweight visualizations to make otherwise difficult-to-interpret numbers easier to assess. These practices helped expose specific triggers, but were often indirect, manual, and easy to miss. P3 and P6 linked delayed detection to lost productivity, with P3 describing it as ``\textit{wasting my agent's time},'' suggesting a need to surface actionable changes sooner.

Across tasks, participants identified four signals that most often triggered intervention. A \textit{completion} signal indicated that a session was ready to move to the next task.
\textit{Requests for clarification} and \textit{errors or bugs} required immediate intervention, yet were also the signals participants most reliably missed: ``\textit{when I come across [an]other session it's stuck there}'' (P1). Finally, an \textit{unexpected delay}—such as a deployment that should have taken two or three minutes but continued past ten—prompted participants to inspect the session and determine whether intervention was necessary.

Thus, triage involved turning concurrent activity into an actionable queue: identifying or constructing signals of completion, blockage, or abnormal progress, then prioritizing interventions.

\section{\pp: A Design Probe for the PILOT Framework}

\begin{figure}[!b]

\definecolor{planningbg}{HTML}{FFFAF2}
\definecolor{loggingbg}{HTML}{F2FCF4}
\definecolor{dashboardbg}{HTML}{EDF9FF}

\DeclareRobustCommand{\componentbox}[2]{%
  \begingroup
  \setlength{\fboxsep}{1pt}%
  \smash{\colorbox{#1}{#2}}%
  \endgroup
}
    \centering
      \vspace{-15px}
    \includegraphics[width=\linewidth]{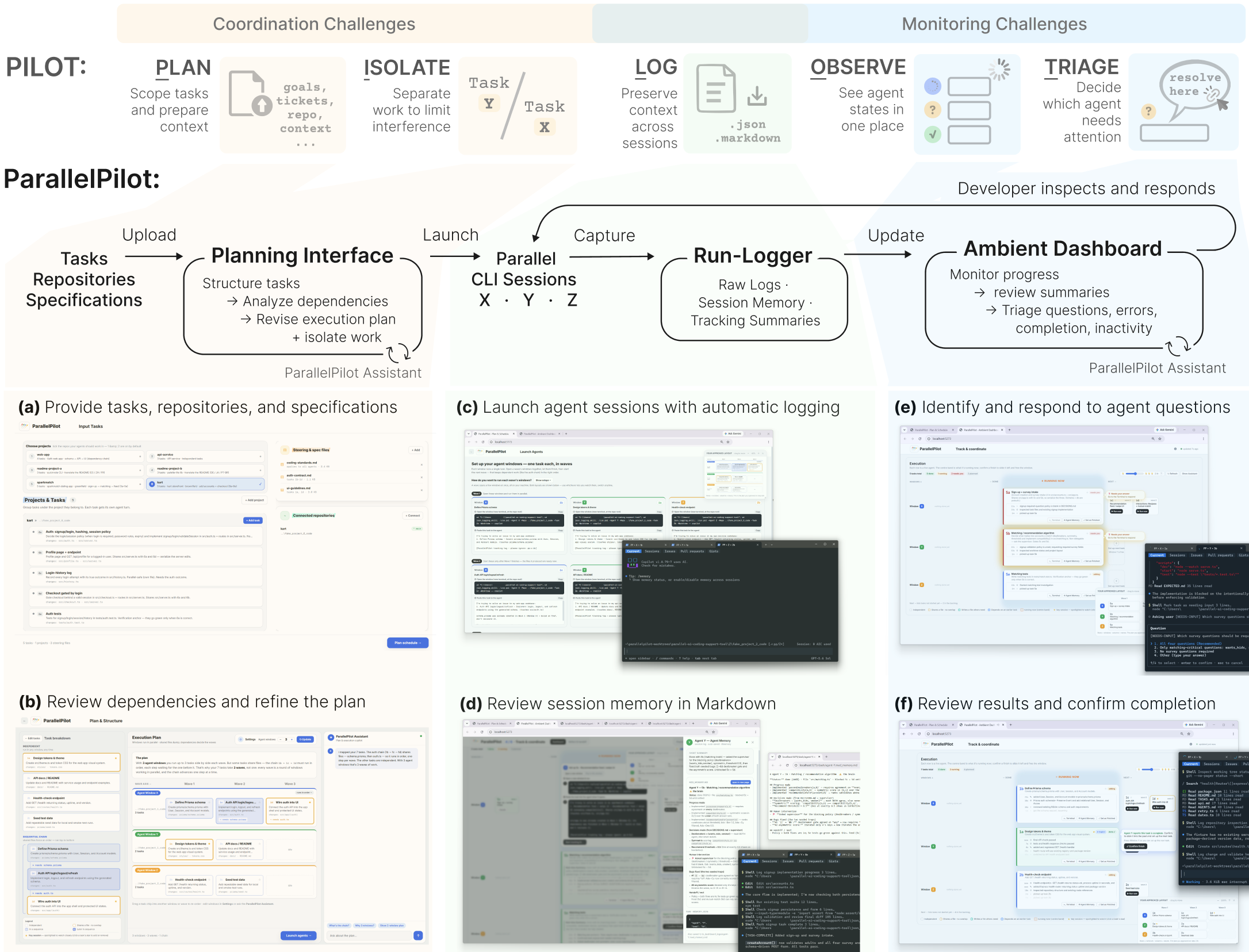}
  
   \caption{\textbf{\pp{} instantiates PILOT through three integrated components:}
a \componentbox{planningbg}{planning interface} for task preparation and dependency-aware scheduling (a--b),
a \componentbox{loggingbg}{run-logger} for automatic capture and session memory (c--d),
and an \componentbox{dashboardbg}{ambient dashboard} for monitoring progress and triaging agent questions and completion (e--f).
The interface thumbnails map these components to successive stages of the workflow; enlarged views of (b) and (e) appear in the system walkthrough.}
\Description{System workflow and six interface screenshots. On the left, tasks, repositories, and specifications feed into the Planning Interface, mapped to Plan and Isolate. Developers refine the plan and launch parallel CLI sessions. The Run-Logger, mapped to Log, captures raw records, session memory, and tracking summaries, which update the Ambient Dashboard, mapped to Observe and Triage. A return arrow shows developers inspecting cues and responding in the CLI sessions. Coordination and monitoring brackets overlap around the Run-Logger. On the right, six screenshots illustrate providing task context, refining the execution plan, launching sessions with automatic logging, reviewing Markdown memory, responding to agent questions, and verifying completion.}
    \label{fig:parallelpilot_overview}
    \vspace{-20px}
\end{figure}

We built \pp{} as a design probe that instantiates the PILOT framework. It is a lightweight web application that runs alongside developers' existing coding CLIs, editors, and workspaces, supporting the {\textit{coordination} and \textit{monitoring}} of parallel AI coding.
There are three integrated components: a planning interface that externalizes dependencies and structures execution (\textbf{Plan, Isolate}); a run-logger that connects agent sessions to tasks and preserves context (\textbf{Log}); and an ambient dashboard that makes agent status glanceable and highlights intervention cues (\textbf{Observe, Triage}). 
Figure~\ref{fig:parallelpilot_overview} maps the three components to the overall workflow, with interface thumbnails providing visual orientation. The walkthrough below presents enlarged views of planning and triage (Figures~\ref{fig:parallelpilot_plan} and~\ref{fig:parallelpilot_triage}).

\subsection{System Design}

We illustrate the design through Amelia, a mid-level software developer persona who has tried parallel AI coding but found it difficult to track overlapping edits and recover context between sessions. 
She is working on two projects: 1) an internal support portal requiring
 \ticket{1a}{Fix ticket pagination},
\ticket{1b}{Add CSV export}, and an \ticket{1c}{improved empty-state} component;
and 2)  a separate data-import CLI requiring
\ticket{2a}{Add CLI retry handling} and
\ticket{2b}{Add retry regression tests}.
The walkthrough follows her from planning and launching agent sessions
to monitoring progress, responding to questions, and returning to
interrupted work.

\subsubsection{\textbf{Planning Interface: Structuring and Isolating Work}}

\begin{figure}[!b]
    \centering
      \vspace{-10px}
    \includegraphics[width=.96\linewidth]{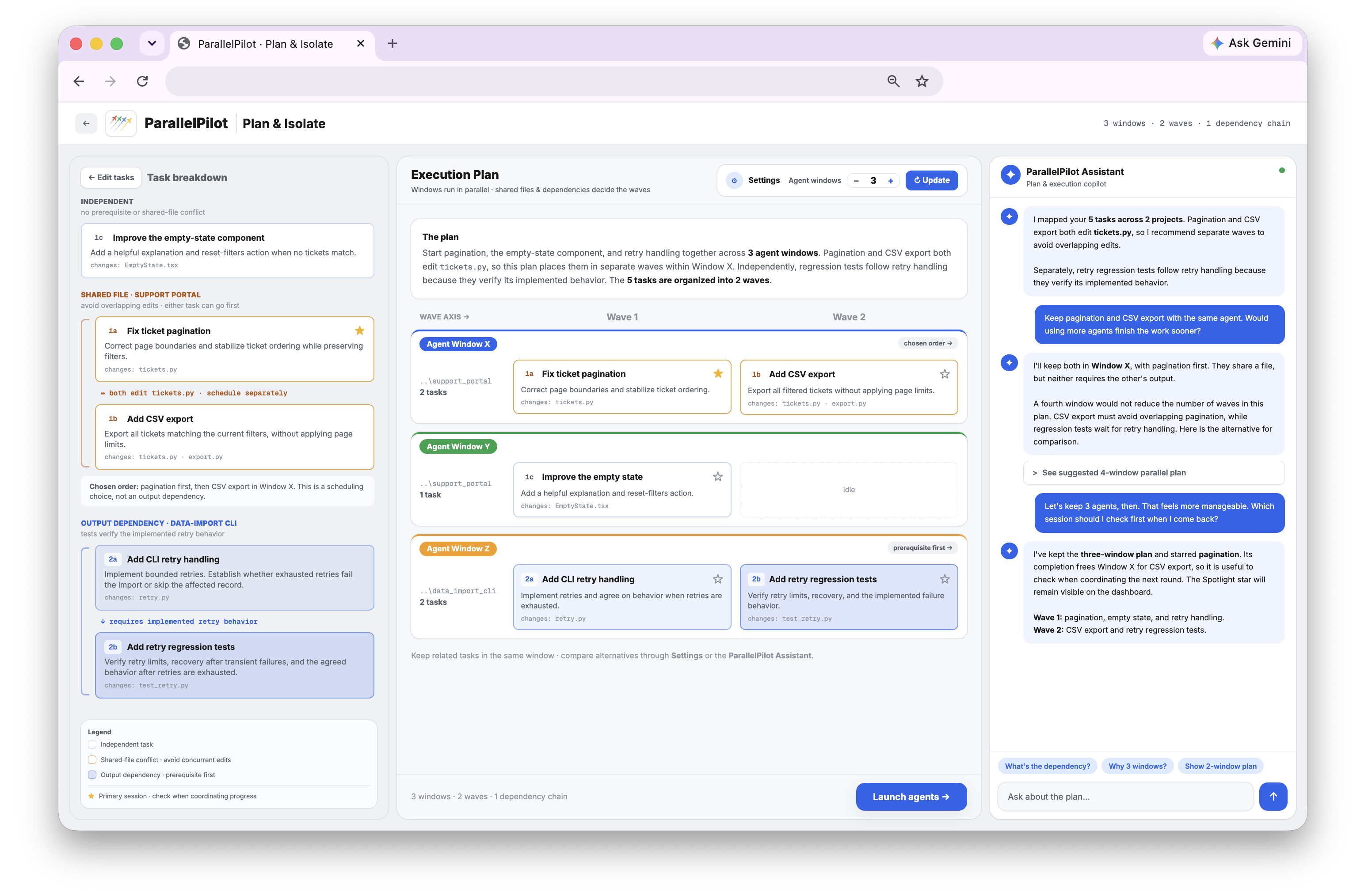}
    \vspace{-15px}
    \caption{\textbf{\pp{} supports dependency-aware plan structuring and iteration.} The task analysis (left) distinguishes independent work, shared-file conflicts, and output dependencies. The execution
timeline (center) organizes five tasks across three agent windows
and two waves, with pagination starred for follow-up.
In the assistant panel (right), Amelia compares a four-window
alternative and retains three windows to keep supervision
manageable.}
\Description{The planning interface has three columns. The left column distinguishes an independent empty-state task, a shared-file conflict between pagination and CSV export in tickets.py, and an output dependency from retry handling to regression tests. The center schedules five tasks across three agent windows and two waves: Window X runs pagination followed by CSV export; Window Y runs the empty-state task; and Window Z runs retry handling followed by regression tests. Pagination is marked with a gold Spotlight star. The right column shows Amelia discussing agent assignments with the assistant, comparing a four-window alternative, and retaining three windows.}
    \label{fig:parallelpilot_plan}
    \vspace{-20px}
\end{figure}

Amelia begins by preparing and building her execution plan
(Figure~\ref{fig:parallelpilot_overview}a--b). She pastes her tickets into the planning interface and attaches the relevant codebases and specification files from her issue-driven engineering workflow. She browses the available \texttt{\small AGENTS.md} files in \pp{} for suitable agent instructions. She then clicks \button{Analyze \& Plan}.

The resulting planning view, shown in detail in
Figure~\ref{fig:parallelpilot_plan}, presents a task-relationship graph on the left, an execution timeline in the center,
and the ParallelPilot Assistant chatbot on the right. The analysis flags that \ticket{1a}{pagination} and \ticket{1b}{CSV export} both modify \texttt{\small tickets.py}, recommending that they run sequentially
to avoid overlapping edits. This surprises Amelia, who has not worked with these files before. Separately, the graph identifies an output dependency: \ticket{2b}{regression tests} must verify the retry limits and failure behavior established during \ticket{2a}{retry implementation}.

The timeline proposes three agent windows and two \textit{waves}:
planning rounds that indicate task order, rather than requiring
all tasks to start or finish together. \ticket{1a}{}, \ticket{1c}{}, and \ticket{2a}{} can proceed in parallel; \ticket{1b}{} and  \ticket{2b}{} follow in their respective windows once
the preceding work is reviewed and confirmed. A task in the
second wave need not wait for unrelated work in other windows.
Amelia asks the assistant, ``\textit{Keep
\ticket{1a}{pagination} and \ticket{1b}{CSV export} with the same
agent. Would using more agents finish the work sooner?}''
The assistant offers a four-window alternative, explaining that
another window would not reduce the number of waves under the
current task breakdown and constraints. Amelia retains three
agent windows to keep supervision manageable. The assistant
marks \ticket{1a}{} with a Spotlight star because its
completion frees that window for \ticket{1b}{},
making it a useful session to check when coordinating the next
round. Satisfied, she accepts the plan and clicks
\button{Launch Agents $\rightarrow$} to open the per-agent
setup instructions. 

Externalizing dependencies and separating potentially conflicting work supports \textbf{Plan} and \textbf{Isolate}, while keeping model-generated recommendations open to user review and revision. Implementation details for dependency analysis, plan revision, and isolated session setup appear in Appendix~\ref{app:planning-implementation}.


\subsubsection{\textbf{Run-Logger: Preserving Session Context}}

Amelia next launches the sessions with automatic logging, which preserves their activity for later review
(Figure~\ref{fig:parallelpilot_overview}c). On the setup page, she finds a terminal command and a jumpstart prompt for each first-round agent. She runs each command to launch a Copilot CLI session, then pastes the corresponding prompt to link it to its assigned task and window ID. Three sessions begin working on \ticket{1a}{}, \ticket{1c}{}, and \ticket{2a}{}. Amelia keeps the terminals on her main monitor and the project tickets on her two extended monitors. 
As the agents work, the run-logger automatically preserves their activity and conversations in three layers: raw records, mid-length session memory, and brief tracking summaries. Amelia uses the tracking summaries to monitor progress at a glance. When a ticket needs closer attention, she opens its session memory to review changes, decisions, and reported results, cross-checking the CLI output as needed rather than reconstructing progress entirely from logs (Figure~\ref{fig:parallelpilot_overview}d). Raw records remain available in her local log folder for detailed inspection.  

By combining detailed records with progressively shorter views, the run-logger supports monitoring and context recovery, instantiating the \textbf{Log} pillar.  Implementation details for conversation capture, local persistence, and memory generation appear in Appendix~\ref{app:logger-implementation}.

\begin{figure}[!b]
    \centering
        \vspace{-5px}
    \includegraphics[width=.95\linewidth]{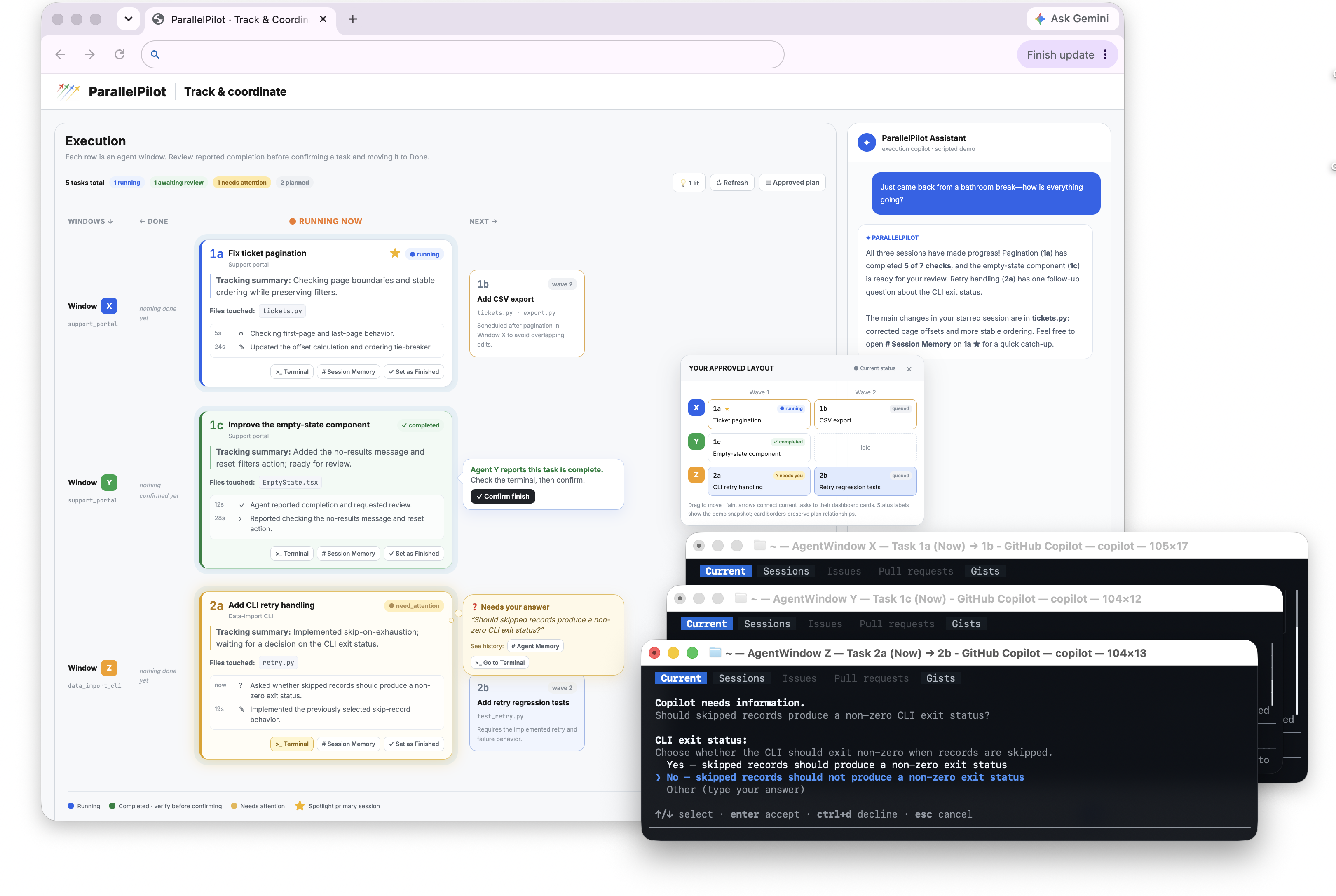}
    \vspace{-10px}
 \caption{\textbf{\pp{} combines glanceable status, completion and attention cues, session memory, and terminal access to support monitoring and triage.}
After her bathroom break, Amelia catches up on parallel progress:
\protect\ticket{1a}{} remains running (blue), \protect\ticket{1c}{} reports
completion pending her review (green), and \protect\ticket{2a}{} needs
her input on the CLI exit status for skipped records (yellow).
The dashboard surfaces completion and attention cues alongside
session memory and a progress summary, while direct terminal
access lets her review results and answer the agent in its
CLI session.}
\Description{The dashboard arranges agent windows in rows with Done, Running Now, and Next columns. Window X shows starred pagination task 1a running in blue, with CSV export queued. Window Y shows empty-state task 1c completed in green, alongside a prompt to review the result and confirm completion. Window Z shows retry task 2a highlighted in yellow, with a thought bubble asking whether skipped records should produce a non-zero CLI exit status. Regression tests remain queued. Cards include activity summaries, files touched, and terminal and session-memory controls. On the right, the assistant summarizes progress after Amelia's bathroom break. A floating approved-plan panel shows task assignments and statuses. Three overlapping terminal windows appear in the foreground; Window Z displays the same exit-status question with response options.}
    \label{fig:parallelpilot_triage}
    \vspace{-25px}
\end{figure}

\subsubsection{\textbf{Ambient Dashboard: Monitoring and Triage}}

During execution, the ambient dashboard helps Amelia track progress,
respond to agent questions, and review reported completion.
Figure~\ref{fig:parallelpilot_triage} provides a detailed view of
these monitoring and triage interactions, introduced in
Figure~\ref{fig:parallelpilot_overview}e--f.
The dashboard opens automatically as Amelia starts the sessions, mapping the approved plan into \textit{Done}, \textit{Running Now}, and \textit{Next} columns. Each card represents a task, with rows corresponding to agent windows and columns indicating execution stage. Active cards show a blue \texttt{\small ``running''} status, files touched, and a one-line activity summary generated from logs. 
About a minute later, just as Amelia is halfway out of her chair
for a bathroom break, the \ticket{2a}{CLI retry card} switches to
a yellow \texttt{\small ``needs\_attention''} status and begins
blinking. A thought bubble surfaces the agent's question about
whether to fail the import or skip a record after exhausting retries.
Glad to catch it before stepping away, Amelia clicks
\button{>\_Terminal} and answers directly in the corresponding
agent session. The card returns to blue \texttt{\small ``running''}
as work resumes. Amelia finds this familiar and comfortable:
\pp{} provides an ambient coordination layer while she stays
in control through direct interaction with the CLI agent. After her break, Amelia checks \button{\# Session Memory} for the
Spotlight-starred \ticket{1a}{} to catch up on its progress.
Meanwhile, \ticket{1c}{} displays a green
\texttt{\small ``completed''} status and prompts her to inspect
the result in the terminal. The \ticket{2a}{CLI retry card} also
raises a follow-up question: ``\textit{Should skipped records
produce a non-zero CLI exit status?}'' These completion and
attention cues appear together in
Figure~\ref{fig:parallelpilot_triage}. Amelia reviews
\ticket{1c}{}'s changes and confirms it as done, then returns to
\ticket{2a}{}'s terminal to clarify the exit-status behavior and
let it continue. Once \ticket{1a}{} completes, she reviews and
confirms its result and starts \ticket{1b}{}.
By the end of the run, Amelia feels she has balanced parallel
progress with manageable supervision. She has spent less effort
coordinating sessions and reconstructing context, while the saved
Markdown summaries preserve a record of changes and decisions
for future reference. 

These features instantiate \textbf{Observe} and \textbf{Triage}, enabling at-a-glance monitoring and targeted attention to sessions with prolonged inactivity, errors, input requests, or completion.
Implementation details for dashboard updates, intervention detection, and completion verification appear in Appendix~\ref{app:dashboard-implementation}.


\subsection{System Implementation}
\label{sec:system-implementation}

\pp{} is a local-first web application that operates alongside
existing coding CLIs.
Its JavaScript/HTML interfaces use Preact and Tailwind CSS,
with lightweight Node.js backends and PowerShell scripts for
startup and session launch. Session records and generated
artifacts are stored on the local filesystem, while
model-assisted processing uses remote APIs.
{The planning interface} uses chained calls to
\texttt{\small gpt-5.6-sol} with medium reasoning, balancing dependency-analysis quality with latency, to
propose editable, dependency-aware assignments and generate
launch instructions for agents working in isolated repository
directories.
{The run-logger} captures Copilot conversations through read-only
SQLite polling and preserves raw records alongside session
memory and tracking summaries generated by
\texttt{\small gpt-5-mini} for speed.
{The ambient dashboard} refreshes from local
records approximately every 2.5 seconds and uses
\texttt{\small gpt-5.6-sol} with medium reasoning to surface potential intervention
needs.
Across these components, developers retain direct access through
their CLI sessions; \pp{} provides the planning, context
preservation, and monitoring layer around them.
Appendix~\ref{app:system-implementation} provides the full
implementation details.

\section{\pp User Study}

\subsection{Overview}

To explore how developers manage the demands of supervising parallel AI coding sessions, we conducted an 80-minute, in-person, counterbalanced within-subjects design probe with 16 developers. The study probes whether and how the PILOT-based design supports parallel AI coding in practice. We examine its effects on objective task outcomes and subjective user experience, as well as how it supports developers' coordination and monitoring across concurrent sessions. Finally, we investigate whether these forms of support translate into greater perceived control, more successful interventions, and increased trust. We operationalized these goals through three RQs and hypotheses (H):



\begin{itemize}

\item \textbf{RQ1 (Overall Effectiveness \& User Experience):}
Does \pp improve task completion\htag{H1a},
throughput\htag{H1b}, supervision capacity\htag{H1c}, and user experience\htag{H1d}?

\vspace{3px}
\item \textbf{RQ2 (Coordination \& Monitoring Support):}
How does \pp support dependency understanding\htag{H2a},
planning and task assignment\htag{H2b}, monitoring effort\htag{H2c},
and awareness of agent states and progress\htag{H2d}?

\vspace{3px}
\item \textbf{RQ3 (Control \& Intervention Dynamics):}
Does \pp support greater perceived control and intervention effectiveness\htag{H3a},
and trust and delegation confidence\htag{H3b}?

\end{itemize}

\subsection{Participants}

We recruited 16 participants (U1--U16) from an internal company research and engineering population through an email invitation. Recruitment targeted developers who regularly use AI coding assistants and have some experience with parallel AI coding workflows. Of the 16 participants (10 women, 6 men), 13 were in the 25--34 age range and three were in the 18--24 range. Fourteen participants reported substantial coding experience: 10 had 4--6 years and 4 had 7--10 years. Fifteen participants reported using AI coding assistants at least once per day, with 13 of 16 reporting use more than three times per day. All had prior parallel AI coding experience, but 15 rated themselves as Beginners or Somewhat Experienced, indicating that they were just getting started or still developing their approach.

Pre-study survey responses ($N=16$) highlighted the challenges identified in the formative study. Eight participants reported difficulty decomposing tasks, and 14 reported experiences with overlapping or conflicting edits. Eleven found concurrent sessions difficult to track, 11 missed agent status changes, and 14 found session switching mentally taxing. Eleven spent effort reconstructing
progress, 10 relied on session logs, 11 struggled to recall prior
accomplishments, and 15 had difficulty reconstructing agents'
decisions. Ten were unsure when to
intervene. These responses were consistent with the coordination
and monitoring needs that \pp{} was designed to address.

\subsection{Study Design and Procedures}
Each participant completed an in-person, 80-minute session, using \pp{} for one project and the no-\pp baseline for the other. Participants were randomly assigned to four groups, counterbalancing tool order and tool-project assignments. The projects, Sparkmatch and Kart, are described in the following subsection.
This counterbalanced design let participants serve as their own control while mitigating order and task-difficulty effects. See Appendix~\ref{app:method-details} for group procedures and Appendix~\ref{app:study-checks-order} for the session-order analysis.

Each 80-minute session began with informed consent and an orientation to the study and apparatus. Participants used a researcher-provided laptop and two external monitors, freely arranging windows across one, two, or three screens. See Appendix~\ref{app:apparatus-details} for apparatus and configuration details, and Appendix~\ref{app:user-study-setup-details} for photos of how participants work with \pp{} across screens. Participants then proceeded through two counterbalanced coding blocks. Each of the two 30- to 35-minute coding blocks began with a 5-minute task introduction, in which the study administrator introduced the seed project, the tickets on a mock GitHub issue page, and walked through the seed codebase in VS Code. Participants had 20 minutes to complete the assigned coding tasks. In the \pp condition, participants additionally watched a 5-minute onboarding video introducing the PILOT framework and \pp, followed by a brief trial setup with the administrator.
In the baseline condition, participants used GitHub Copilot CLI or Chat on the researcher-provided workstation. They could plan with the coding assistant, open multiple sessions, create worktrees, and maintain notes. 
Each block ended with a 5-minute post-condition survey assessing UX, coordination, and monitoring load, and perceived control. 
Sessions concluded with a post-study comparative survey and a 10-minute semi-structured interview reflecting on their experience across conditions. 
Participants also completed a pre-study survey covering demographics and prior experience with (parallel) AI coding.
All participants provided informed consent to participate and to have the interview, screen, and code recorded for analysis. They were compensated 50 USD, and the study protocol was approved by the internal ethics committee.  See Appendices~\ref{app:pre-survey}, \ref{app:post-condition-survey}, \ref{app:comparative-survey}, and \ref{app:interview} for the pre-survey, post-condition survey, comparative survey, and interview questions.

\subsection{Task Design}
Participants worked on Sparkmatch, a minimally implemented dating application, and Kart, a partially implemented social shopping application. Each project included six tickets with explicit requirements and acceptance criteria on a mock GitHub issue page. Each project included two tasks requiring changes to the same file, one dependency chain of two tasks requiring cross-task integration, and two other independently implementable tasks. One task in each project also required a product or policy decision. These features created opportunities to coordinate parallel work and resolve design choices with coding agents.  To focus on implementation coordination and monitoring, neither condition required participants to manage pull requests, review code, or merge changes.  Two developers who did not participate in the study pilot-tested the projects to assess comparable difficulty and completion times. Detailed task descriptions appear in Appendix~\ref{app:task-details}, and perceived-difficulty checks appear in Appendix~\ref{app:study-checks-tasks}.

\subsection{Data Collection and Analysis}
Our dataset comprised four surveys per participant (pre-study, two post-condition, and post-study comparison), interview recordings and transcripts, screen recordings, and system logs. We summarized survey responses descriptively and compared conditions using paired $t$-tests with Benjamini--Hochberg correction~\cite{benjamini_Hochberg} for multiple comparisons. Results were considered statistically significant when unadjusted $p<.05$ and BH-adjusted $q<.05$. 
We used inductive
affinity diagramming~\cite{affinityDiagramming, thematicAnalysis} across transcripts and researcher notes,
refining themes through feedback from all co-authors.

\newcommand{\hypidPaper}[1]{{\small\textbf{\texttt{#1}}}}

\section{\pp User Study Results}
\label{sec:results}

\normalsize
\subsection{\textbf{RQ1 (Overall Effectiveness \& User Experience):} Does \pp improve task completion\htag{H1a}, throughput\htag{H1b}, supervision capacity\htag{H1c}, and user experience\htag{H1d}?}

\label{sec:results-rq1}


\summary{\small Compared with the baseline, objective logs showed that \pp{} increased tickets completed per minute by 63\% and total completion by 1.50 tickets. Participants scaled up parallel work by about 1 agent with \pp{} and anticipated supervising 1.44 more agents in future use than without it. Survey ratings showed significantly higher perceived efficiency (+1.69/7 points), better UX, and less perceived conflict. Table~\ref{tab:hypotheses-results} provides evidence supporting \hypidPaper{H1a}--\hypidPaper{H1d}.}



\newpage
\vspace*{-30px}

{\footnotesize

\definecolor{supportedrow}{RGB}{235,247,237}
\newcommand{\supportedrow}{\rowcolor{supportedrow}}
\DeclareRobustCommand{\green}[1]{%
  \begingroup
  \setlength{\fboxsep}{1pt}%
  \smash{\colorbox{supportedrow}{#1}}%
  \endgroup
}

\providecommand{\hypgap}{}
\renewcommand{\hypgap}{\addlinespace[3pt]}

\newcommand{\hypsb}{\fontseries{sb}\selectfont}
\renewcommand{\hyp}[3][]{%
  \raggedright
  \hspace{-18pt}%
  \if\relax\detokenize{#2}\relax
    \colorbox{hypidbg}{%
      \begin{tikzpicture}[baseline=-0.5ex]
        \path[use as bounding box]
          (0em,-0.25em) rectangle (1.25em,0.25em);
        \begin{scope}[overlay]
          \if\relax\detokenize{#1}\relax
            \draw[line width=0.45pt, gray!60]
              (0.45em,0.85em) -- (0.45em,-0.85em);
          \else
            \draw[line width=0.45pt, gray!60]
              (0.45em,0.85em) -- (0.45em,-0.25em);
          \fi
          \draw[line width=0.45pt, gray!60]
            (0.45em,-0.25em) -- (1.15em,-0.25em);
        \end{scope}
      \end{tikzpicture}%
    }%
    \hspace{2px}%
    {\footnotesize\hypsb #3}%
  \else
    \colorbox{hypidbg}{%
      \rule{0pt}{0.2em}%
      \textbf{\scriptsize\texttt{#2}}%
    }%
    \hspace{0.25em}%
    {\footnotesize\hypsb #3}%
  \fi
}

\tikzset{deltabar/.style={baseline=-0.6ex,x=1.2cm,y=1cm}}
\renewcommand{\posbar}[2]{%
  \begin{tikzpicture}[deltabar]
    \draw[gray!35,line width=0.4pt] (-0.42,0) -- (0.42,0);
    \draw[gray!55,line width=0.4pt] (0,-0.045) -- (0,0.045);
    \fill[ppgreen!70] (0,-0.035) rectangle (#1,0.035);
  \end{tikzpicture}\,\hspace{2px}#2%
}
\renewcommand{\negbar}[2]{%
  \begin{tikzpicture}[deltabar]
    \draw[gray!35,line width=0.4pt] (-0.42,0) -- (0.42,0);
    \draw[gray!55,line width=0.4pt] (0,-0.045) -- (0,0.045);
    \fill[ppgreen!70] (-#1,-0.035) rectangle (0,0.035);
  \end{tikzpicture}\,\hspace{2px}#2%
}
\renewcommand{\nobar}[1]{%
  \begin{tikzpicture}[deltabar]
    \draw[gray!25,line width=0.4pt] (-0.42,0) -- (0.42,0);
    \draw[gray!45,line width=0.4pt] (0,-0.045) -- (0,0.045);
  \end{tikzpicture}\,\hspace{2px}#1%
}

\setlength{\tabcolsep}{3pt}
\renewcommand{\arraystretch}{1.12}
\begin{xltabular}{\linewidth}{@{}
    l
    l
    >{\centering\arraybackslash\hsize=0.70\hsize}X   
    >{\centering\arraybackslash\hsize=1.20\hsize}X   
    >{\raggedright\arraybackslash\hsize=1.60\hsize}X 
    >{\raggedleft\arraybackslash\hsize=0.60\hsize}X  
    >{\raggedleft\arraybackslash\hsize=0.80\hsize}X  
@{}}
\\
\caption{Survey- and log-based results for RQ1, RQ2, and RQ3, comparing parallel AI coding with and without \pp ($N=16$). Means are on a 1--7 scale unless otherwise noted. Paired $p$-values are two-tailed; BH $q$ reports Benjamini--Hochberg-adjusted $p$-values
across all hypothesis tests in the table. For both columns, * indicates $<.05$, ** indicates $<.01$, and *** indicates $<.005$. $\dagger$ indicates borderline BH-adjusted values $.05 \le q < .06$. \green{Green rows indicate measures that provide evidence supporting the corresponding RQ.} Delta bars visualize the direction and approximate magnitude of the difference between conditions. \textit{Italicized} measures are survey items, adapted and abbreviated for space; see Appendix~\ref{app:post-condition-survey} for full wording.}
\label{tab:hypotheses-results}
\Description{Table summarizing survey and behavioral evidence for three research questions comparing parallel AI coding with and without \pp. Rows are grouped by research question and report the measure, Without-\pp mean, With-\pp mean, change, paired p-value, and BH-adjusted q-value. Green rows indicate measures providing evidence in support of the corresponding research question.}
\\
\toprule
\noalign{\vskip 2pt}
\textbf{\small RQs / hypotheses} & \textbf{\small Measure / survey item} &
\textbf{\small \hspace{-10px}Baseline} &
\textbf{\small ParallelPilot} &
\multicolumn{1}{c}{\small \textbf{$\Delta$}} &
\textbf{\small $p$} &
\textbf{\small BH $q$}\\
\noalign{\vskip 2pt}
\midrule
\endfirsthead
\toprule
\textbf{RQs / hypotheses} & \textbf{Measure / survey item} &
\textbf{Baseline} &
\textbf{ParallelPilot} &
\multicolumn{1}{c}{\textbf{$\Delta$}} &
\textbf{$p$} &
\textbf{BH $q$}\\
\midrule
\endhead

\noalign{\vskip 1pt}
\multicolumn{7}{l}{\hspace{-0px}\textbf{{\small RQ1: Overall Effectiveness \& User Experience}}}\\
\noalign{\vskip 3pt}

\supportedrow
\hyp{H1a}{Ticket completion $\uparrow$}
& { Tickets finished (of 6)}
& 4.19
& 5.69
& \posbar{0.32}{+1.50}
& .018*
& .028*\\

\supportedrow
\hyp[last]{}{Full-completion participants $\uparrow$}
& Participants finishing all 6 tickets
& 8/16
& 14/16
& \posbar{0.34}{+6}
& --
& --\\
\hypgap



\supportedrow
\hyp{H1b}{Ticket Throughput $\uparrow$}
& Tickets per minute
& 0.272
& 0.445
& \posbar{0.40}{+63\%}
& .003***
& .007**\\

\supportedrow
\hyp{}{Task completion time $\downarrow$}
& Time on task (min)
& 16.33
& 13.76
& \negbar{0.24}{-2.57}
& .036*
& .050$\dagger$\\

\supportedrow
\hyp[last]{}{Dual-completer finish time $\downarrow$}
& Dual completers' time on task (min)
& 15.87
& 12.43
& \negbar{0.33}{-3.44}
& .004***
& .007**\\
\hypgap


\supportedrow
\hyp{H1c}{Observed peak concurrency $\uparrow$}
& Max concurrent agents used
& 2.31
& 3.25
& \posbar{0.20}{+0.94}
& .002***
& .005**\\
\supportedrow
\hyp[last]{}{Perceived supervision capacity $\uparrow$}
& Agents comfortable supervising in future
& 2.56
& 4.00
& \posbar{0.31}{+1.44}
& .001***
& .002***\\
\hypgap

\supportedrow
\hyp{H1d}{Perceived efficiency $\uparrow$}
& \textit{I felt efficient working this way.}
& 3.75
& 5.44
& \posbar{0.37}{+1.69}
& $<$.001***
& $<$.001***\\

\supportedrow
\hyp{}{Reuse intent $\uparrow$}
& \textit{I'd work this way again on real tasks.}
& 3.81
& 5.63
& \posbar{0.40}{+1.81}
& $<$.001***
& $<$.001***\\

\supportedrow
\hyp{}{Ease of use $\uparrow$}
& \textit{The setup was easy to use.}
& 4.44
& 5.56
& \posbar{0.26}{+1.13}
& .014*
& .015*\\

\supportedrow
\hyp{}{Needs fit $\uparrow$}
& \textit{It met my needs for supervising agents.}
& 3.75
& 5.31
& \posbar{0.34}{+1.56}
& $<$.001***
& $<$.001***\\

\supportedrow
\hyp[last]{}{Workflow preference $\uparrow$}
& \textit{I prefer \pp over my current setup.}
& --
& 14/16
& \nobar{--}
& --
& --\\

\noalign{\vskip 3pt}

\midrule

\noalign{\vskip 2pt}
\multicolumn{7}{l}{\hspace{-0px}\textbf{{\small RQ2: Coordination Attention}} -- Task structuring (Plan, Isolate)}\\
\noalign{\vskip 3pt}

\supportedrow
\hyp{H2a}{Dependency understanding $\uparrow$}
& \textit{I understood dependencies between tasks.}
& 3.44
& 5.50
& \posbar{0.42}{+2.06}
& $<$.001***
& $<$.001***\\

\supportedrow
\hyp{}{Perceived agent overlap $\downarrow$}
& \textit{Agents worked on overlapping code.}
& 3.06
& 2.13
& \negbar{0.22}{-0.94}
& .038*
& .051$\dagger$\\

\supportedrow
\hyp[last]{}{Isolation usefulness $\uparrow$}
& \textit{Usefulness: dependency \& isolation support}
& --
& 6.33
& \nobar{--}
& --
& --\\
\hypgap

\supportedrow
\hyp{H2b}{Plan clarity $\uparrow$}
& \textit{I had a clear plan to parallelize the work.}
& 3.63
& 5.00
& \posbar{0.31}{+1.38}
& .007**
& .012*\\

\supportedrow
\hyp{}{Task-agent assignment fit $\uparrow$}
& \textit{My assignments fit what each task needed.}
& 3.81
& 4.56
& \posbar{0.18}{+0.75}
& .013*
& .034*\\

\supportedrow
\hyp{}{Planning app usefulness $\uparrow$}
& \textit{Usefulness: plan \& dependency view}
& --
& 6.63
& \nobar{--}
& --
& --\\

\supportedrow
\hyp[last]{}{Logs usefulness $\uparrow$}
& \textit{Usefulness: session logs}
& --
& 5.09
& \nobar{--}
& --
& --\\


\noalign{\vskip 3pt}
\multicolumn{7}{l}{\hspace{0px}\textbf{{\small RQ2: Monitoring Attention}} -- Awareness \& recall (Log, Observe, Triage)}\\
\noalign{\vskip 3pt}

\supportedrow
\hyp{H2c}{Tracking effort $\downarrow$}
& \textit{Tracking every agent took mental effort.}
& 4.63
& 2.75
& \negbar{0.39}{-1.88}
& $<$.001***
& $<$.001***\\

\supportedrow
\hyp{}{Bookkeeping effort $\downarrow$}
& \textit{I needed a lot of manual bookkeeping.}
& 4.63
& 2.50
& \negbar{0.42}{-2.13}
& $<$.001***
& $<$.001***\\

\supportedrow
\hyp{}{Switching disruption $\downarrow$}
& \textit{Switching sessions disrupted my focus.}
& 4.50
& 2.69
& \negbar{0.38}{-1.81}
& $<$.001***
& .002***\\

\supportedrow
\hyp[last]{}{Context-switching $\downarrow$}
& \textit{I was context-switching, not progressing.}
& 3.81
& 2.25
& \negbar{0.34}{-1.56}
& $<$.001***
& .002***\\
\hypgap

\supportedrow
\hyp{H2d}{Intervention cue awareness $\uparrow$}
& \textit{I noticed when agents finished or needed me.}
& 3.63
& 6.19
& \posbar{0.42}{+2.56}
& $<$.001***
& $<$.001***\\

\supportedrow
\hyp{}{Individual agent awareness $\uparrow$}
& \textit{I knew what each agent was working on.}
& 4.00
& 5.06
& \posbar{0.25}{+1.06}
& .044*
& .057$\dagger$\\

\supportedrow
\hyp{}{Overall progress awareness $\uparrow$}
& \textit{I had a clear picture of overall progress.}
& 3.44
& 6.06
& \posbar{0.42}{+2.63}
& $<$.001***
& $<$.001***\\

\supportedrow
\hyp[last]{}{Dashboard usefulness $\uparrow$}
& \textit{Usefulness: live dashboard}
& --
& 6.88
& \nobar{--}
& --
& --\\

\noalign{\vskip 3pt}
\midrule


\noalign{\vskip 2pt}
\multicolumn{7}{l}{\hspace{0px}\textbf{{\small RQ3: Control \& Intervention Dynamics}}}\\
\noalign{\vskip 3pt}

\hyp{H3a}{Intervention success $\uparrow$}
& \textit{My interventions redirected the agent.}
& 3.94
& 4.13
& \posbar{0.05}{+0.19}
& .603
& .619\\

\hyp{}{Intervention confidence $\uparrow$}
& \textit{I felt confident steering an agent.}
& 3.69
& 3.81
& \posbar{0.03}{+0.13}
& .556
& .646\\

\hyp[last]{}{Perceived control $\uparrow$}
& \textit{I felt in control of the agents.}
& 3.94
& 4.31
& \posbar{0.09}{+0.38}
& .414
& .416\\
\hypgap

\hyp{H3b}{Output trust $\uparrow$}
& \textit{I trusted the output.}
& 4.50
& 4.69
& \posbar{0.05}{+0.19}
& .261
& .416\\

\hyp[last]{}{Delegation confidence $\uparrow$}
& \textit{I felt confident delegating without watching.}
& 4.06
& 4.63
& \posbar{0.14}{+0.56}
& .108
& .128\\
\noalign{\vskip 3pt}
\bottomrule
\end{xltabular}
}

\newpage

\subsubsection{\textbf{\hypidPaper{[H1a \& H1b]} Higher completion and faster work from logs.}}
Objective behavioral data show that \pp enabled participants to complete more work in less time. 
 We used a lightweight implementation check, cross-referencing
session recordings with the resulting code to verify that reported ticket completions corresponded to implemented
changes.
With \pp, participants completed an average of 1.50 more tickets than without it (5.69 vs.\ 4.19 of 6 tickets, $q = 0.028$). They also spent 2.57 fewer minutes on task (13.76 vs.\ 16.33 min) and achieved 63\% higher throughput, completing 0.445 vs.\ 0.272 tickets per minute ($q=0.007$). Completion rates increased substantially: 14 of 16 participants completed all six tickets with \pp, compared with 8 of 16 without it. Among participants who completed all six tickets in both conditions, 
those using
\pp finished 3.44 minutes faster on average (12.43 vs.\ 15.87 min, a reduction of
roughly 22\%). The split between coordination (planning and isolating) and later
monitoring was essentially unchanged---25\%/75\% with \pp against 27\%/73\%
without---so the gain came from compressing both phases rather than from shifting
effort between them  (see details in Figure~\ref{fig:phase-time} in
Appendix~\ref{app:phase-time}).


\subsubsection{\textbf{\hypidPaper{[H1c \& H1d]} Improved perceived efficiency, parallel supervision capacity, and UX}}
Self-reported measures closely aligned with the objective performance improvements. Participants rated their own efficiency 1.69 points higher with \pp, increasing from 3.75 to 5.44 on a 7-point scale ($q < 0.001$). 
Participants reported that \pp was easier to use, better suited to their needs, and more useful for future parallel AI coding, with all corresponding comparisons statistically significant. Participants described \pp as ``\textit{much easier than [my setup]}'' (U6), ``\textit{convenient and easier to proceed... very easy to use. I don't have to do a lot of manual work; it's more efficient}'' (U1). 

With \pp{}, participants used about 1 more agent in parallel than in the baseline (3.25 vs.\ 2.31 at peak use). 
Participants, who generally had limited parallel-coding experience, also anticipated comfortably supervising 1.44 more agents with \pp{} in future use ($p=0.001$, $q=0.002$), indicating increases in both observed concurrency and perceived supervision capacity.
Overall, a strong majority (14/16) preferred \pp{} over baseline, with two exceptions: U9 found the experiences ``\textit{still somewhat similar},'' while U16 preferred the baseline, citing setup overhead and worrying that high-level abstractions reduced their control (``\textit{scared of giving up control}'') and engagement with implementation details (``\textit{[I ended up] not thinking as hard}'').  We explore these perspectives further in Section~\ref{sec:results-rq5}.


%

\subsection{\textbf{RQ2 (Coordination \& Monitoring Support)}:  How does \pp support dependency understanding\htag{H2a},  plan clarity\htag{H2b}, monitoring effort\htag{H2c},
and awareness of agent states and progress\htag{H2d}?}

\label{sec:results-rq2}

\summary{\small Survey ratings showed that \pp{} improved dependency understanding (+2.06/7 points) and final plan clarity (+1.38/7), reduced tracking effort (-1.88/7), and increased intervention-cue awareness (+2.56). Table~\ref{tab:hypotheses-results} provides evidence supporting (\hypidPaper{H2a}--\hypidPaper{H2d}). Exploratory correlations tied coordination gains to objective productivity, monitoring gains to perceived efficiency.}

\subsubsection{\textbf{\hypidPaper{[H2a \& H2b]} Improved coordination: clearer decomposition and dependency awareness.}}

\pp substantially helped participants understand relationships among parallel tasks and build clear plans. Participants' ratings of their understanding of task dependencies increased from 3.44 to 5.50 ($q < 0.001$). Participants also reported less perceived overlap and conflict among agents' work from 3.06 to 2.13, thus rating the isolation feature as highly useful (6.33/7). Six of the eight participants who used \pp first reported later applying what they learned to the baseline condition, particularly by asking about file-sharing conflicts and sequencing dependencies between tasks. 
Building upon the beneficial decompositions, participants reported greater clarity in their overall plans, increasing from 3.63 to 5.00 ($q = 0.012$) and a better fit between tasks and assigned agents, increasing from 3.81 to 4.56. Participants then found the whole planning interface useful (6.63/7) and session logging useful for maintaining the information needed to coordinate work across agents and return to the task context later (5.09/7). U4 shared ``\textit{\pp is better at planning parallelization than me [... simply giving] Copilot a context dump [and asking it to execute]}.''

\subsubsection{\textbf{\hypidPaper{[H2c \& H2d]} Improved monitoring: faster status awareness with less tracking effort.}}

\pp also reduced the effort required to monitor ongoing agent activity while improving participants' awareness of what was happening across sessions. Participants reported substantially less effort spent tracking parallel work from 4.63 to 2.75  ($q < 0.001$) and less bookkeeping to keep multiple agents on track from 4.63 to 2.50. U1 highlighted how separating the dashboard from individual CLI windows helped them ``\textit{keep track of overall progress better and manage multiple agents' contexts [\dots], keep them separate but still stay engaged in both.}'' With \pp, participants experienced less disruption when
switching sessions (from 4.50 to 2.69) and less of a constant sense of
context-switching rather than making progress (from 3.81 to 2.25).
The resulting reduction in monitoring effort was accompanied by stronger situational awareness at both the local and global levels. Ratings of noticing when agents finished, stalled,
or required input increased from 3.63 to 6.19  ($q < 0.001$). Awareness of each agent's current work increased
from 4.00 to 5.06, while awareness of overall
progress increased from 3.44 to 6.06  ($q < 0.001$). U10 described seeing all three agents' completion notifications as ``\textit{the most satisfying UX feature},'' because it made their status immediately visible and allowed them to move on without unnecessary waiting. Participants therefore rated the live monitoring dashboard highly
useful (6.88/7).

Exploratory correlations linked perceived coordination gains
to gains in the objective efficiency (\hypidPaper{H1a} ticket completion and \hypidPaper{H1b} throughput), and perceived
monitoring gains to gains in perceived efficiency (\hypidPaper{H1c}).
These associations are preliminary and do not establish
component-specific effects; Appendix~\ref{app:exploratory}
reports the analysis.

\subsection{\textbf{RQ3 (Control \& Intervention Dynamics):} Does 
Does \pp support greater perceived control and intervention effectiveness\htag{H3a},
and trust and delegation confidence\htag{H3b}?}
\label{sec:results-rq5}

\summary{\small Despite improved intervention-cue awareness, intervention outcomes and perceived control did not significantly improve. Participants noted that \pp{} removed bookkeeping and tracking effort that had reinforced their sense of control, while \pp's higher-level view changed how they engaged with implementation details.
}


\subsubsection*{\textbf{\hypidPaper{[H3a \& H3b]}
Participants maintained similar levels of control-related outcomes, and why}}
One of the clearest improvements in RQ2 was participants' ability to detect cues of when agents needed intervention. However, we did not detect corresponding improvements in perceived control and intervention success. They changed little, from 3.94 to 4.31 and from 3.94 to 4.13, respectively. Confidence in their interventions, output trust, and delegation confidence also did not improve significantly. Consistent with these results, when asked which condition made output easier to verify, 9 participants reported no difference, 4 preferred \pp{}, and 3 preferred the baseline.
Participants' interview accounts suggest reasons why easier coordination and monitoring did not necessarily produce greater control or trust:

\paragraph{Coordination and monitoring support complement, rather than replace, evaluation and control.}
Knowing when to intervene does not necessarily mean knowing how to intervene. \pp{} focused on helping developers coordinate parallel tasks, follow progress, and identify agents needing attention. Participants' accounts suggest that these benefits addressed a different need from evaluating an agent's approach or correcting its implementation. For example, identifying a stalled agent provides a cue to inspect its work, but determining whether an active agent is pursuing the right approach may require implementation-level evidence beyond summary-oriented progress reports. Also, \pp's plan-focused execution structure appeared to orient attention toward keeping work moving: identifying idle agents and detecting blockers. These are valuable coordination activities, while assessing whether an active agent is doing the right work can require a different kind of effort. An agent that appears busy or reports completion may still benefit from substantive inspection. When \pp surfaced a plan reference, U3 described ``\textit{a tendency to just keep going forward... less likely to ask follow-up questions in the terminal and deeply inspect}.'' 
 For trust and delegation confidence, participants noted that \pp primarily helped them supervise overall progress rather than evaluate the reliability of individual agents or the capabilities of the underlying AI models. 

\paragraph{Coding assistants distance developers from the code, and \pp may now further shift them toward a managerial perspective.}
\pp's existence as an abstraction layer helped participants manage plans, assignments, and progress across agents. The externalized tracking of \pp reduced bookkeeping effort, while participants described some manual tracking activities as opportunities to reinforce their understanding. U15 and U1 shared that, in their native workflow, every time they context-switch between sessions to check on progress, they reinforce their memory by reconstructing and scrutinizing what each agent is working on and how far it has progressed.
Some participants described reconstructing session state less often because an automatically updated overview and markdown files were available.  U10 described it as ``\textit{giving up my internal representation.}'' U16 shared ``\textit{[with \pp,] I [ended up] not thinking as hard}'' and U2 mentioned ``\textit{I feel like I'm getting dumber}'' in the process. An automatically updated overview, therefore, supported awareness without necessarily preserving the detailed familiarity needed for intervention. Returning to agent terminals, inspecting changes, and asking targeted follow-up questions still required context reconstruction. 
Also, U5 described a shift in focus: ``\textit{With \pp, I will focus less on what each agent is doing but overall progress}.'' U11 similarly drew an analogy to an engineer becoming a manager: their focus shifted \textit{``from hands-on implementation [to...] now high-level goals and roadmap''}, leaving less detailed context for stepping in when an agent encountered problems.

\section{Discussion}


\subsection{Learning from Human Management Practices}

PILOT adapts concepts from human supervisory control literature~\cite{theory_coordination_monitoring2} to motivate two areas of support: \textit{coordination}, encompassing planning and teaching, and \textit{monitoring}, encompassing monitoring and intervention. \pp operationalizes these concepts in an interface for parallel AI coding. In our evaluation, participants found that \pp helped address coordination and monitoring challenges, suggesting the value of adapting human management frameworks to agent supervision. 
Across both studies, six participants further used human management metaphors to describe their mental models in coordinating parallel agents. P6, a senior engineering manager, drew on their experience managing junior team members, describing how they used stand-up updates to assess whether someone’s approach aligned with what they considered an effective way forward. They applied a similar strategy to agent supervision, checking not only whether work was progressing, but also whether the agent was approaching the task appropriately. Senior researcher participants and PhD participants similarly described matching agents to tasks based on their ``background'' and skillsets, such as relevant specification files, much as they would assign research assistants based on their expertise. This suggests that prior supervisory experience can provide strategies for managing parallel AI coding.

Our evaluation suggests that interface support like \pp can help developers with varying levels of familiarity manage concurrent agents. However, tools are only one avenue for supporting this work. A complementary opportunity is to teach supervision strategies explicitly: how to structure parallel tasks, preserve shared context, establish checkpoints, recognize the need for intervention, and allocate attention across sessions. Whether training grounded in human management practices improves agent supervision remains an open empirical question. Future work could examine which practices transfer effectively, which require adaptation, and which introduce misleading expectations about agent capabilities or behavior.
This inquiry also requires examining where the human management analogy breaks down. Human supervision involves social dynamics~\cite{socialdynamics}: relationships, mutual accountability, and professional development. Productivity-oriented agent use therefore raises questions about which social dynamics remain relevant and how they shape developers' expectations and oversight. Rather than treating PILOT as a complete solution, we see it as a starting point for studying how interface support and training can jointly develop effective agent-supervision practices.

\subsection{The Abstraction Tension:
Designing Future Human-AI Tools for  Re-engagement}

\pp{} illustrates the value of an abstraction layer over existing coding agents. Its planning interface externalizes task assignments and dependencies, while logs, status cards, and triage cues bring distributed activity into a shared view. Participants could notice agent needs more reliably, switch sessions less disruptively, and complete more work. At the same time, better visibility did not necessarily provide all of the context needed to evaluate outputs or intervene.

Participants' accounts suggest a tension between maintaining a high-level overview and staying engaged with implementation details. Externalized tracking reduced bookkeeping, and some developers described relying less on their own reconstruction of implementation progress and decisions. Knowing that an agent was finished, blocked, or requesting input provided useful situational awareness, but did not by itself establish whether its code was correct or its approach appropriate. For designers of similar tools, this suggests that abstraction should support not only an overview but also \textit{re-engagement}: a clear path from summarized status to the code, decisions, and evidence needed for judgment.

This distinction matters when evaluating HCI productivity tools more broadly. Such tools often abstract complex workflows and automate selected activities to reduce human effort. Yet reductions in effort can change how users engage with the underlying work. 
This consideration resonates with questions of authenticity and authorship in co-creative tools, where reducing production effort raises questions about what constitutes meaningful human involvement~\cite{authenticity_cocreativeKaty,authenticity_cocreativeAngel,taolongitudinal}. In visual analytics and clinical decision-support tools, reducing information-search and processing overhead should preserve users' ability to interpret evidence, question assumptions, and exercise judgment~\cite{authenticity_Med,feedquac}. Across these settings, the design challenge is not simply to minimize human effort, but to distinguish avoidable overhead from engagement that supports informed decisions. In parallel AI coding, meaningful involvement need not require writing every line, but it does require opportunities to understand, judge, and shape the resulting work.  


Our findings point to early design insights for future coding or adjacent human-AI tools navigating this abstraction tension. First, tools should lower the cost of the inevitable re-engagement moment: when a developer needs to invest brainpower to reconstruct context. The moves in \pp{} are: one-click return to the original CLI and fallback resources; clear lead-in messages and timely notifications, reducing the need to reconstruct context and preventing delayed awareness; and persistent tracking and saving of history and files, which keep disengaged work retrievable and promote a sense of security.
Second, and more fundamentally, how much a tool abstracts away, and thus how much developers disengage, is itself a design decision, not a fixed benefit or side effect of automation. A schema change and a syntax fix warrant different defaults, which is why the abstraction level should be specified per task rather than fixed per tool. But whichever level is chosen, it should remain legible to the developer and it should be revisited over time--as a developer gains history with the system or proficiency with the tasks, the level of detail they actually need may shift, and whether that shift stays visible to them, rather than happening silently, is a genuine alignment question.

Scaffolding people out of as much of the work that does not require them as possible is not a new ambition in HCI. What our findings sharpen is the harder half of that ambition: calibrating how far disengagement can go before it costs judgment, aligning to users' needs and expectations as they shift over usage and time, and designing for re-engagement at the moments that actually call for it, rather than leaving developers to notice on their own.

\section{Limitations and Future Work}

Our work has several limitations. First, our formative study included 14 participants, primarily researchers and research interns. Though they reflect early adopters of parallel AI coding, future work should examine broader industry populations and development constraints, including speed-first MVP development and open-ended research engineering.

Second, \pp is a design probe of our framework, not an exhaustive exploration of its design space. Its dependency-aware plans and textual summaries offer one approach to coordination and monitoring, while its four intervention cues—completion, errors, questions, and being stuck—may not capture the full range of real-world intervention needs. Also, our findings depend on the specific model (\texttt{\small gpt-5-mini} for log summarization and \texttt{\small gpt-5.6-sol} with medium reasoning on all other tasks; access date: mid-August 2026) and the tool configurations used for \pp and GitHub Copilot. Future work should explore alternative ways to abstract, surface, and act on information, and examine their effects on attention, context recovery, and trust, as well as other model families.

Third, our \pp user study involved 16 participants in a counterbalanced within-subjects design using 20-minute blocks and six seed tasks. The short blocks may underestimate context-recovery costs and overestimate developers' ability to retain session state, limiting ecological validity. The fixed set of six seed tasks may also introduce a ceiling effect for completion: participants who completed all six tasks before the block ended could not demonstrate additional throughput. In addition, our evaluation focused on implementation throughput, so participants were not required to conduct systematic code review. Thus, we did not directly measure verification accuracy or implementation-level understanding. Our comparison evaluates ParallelPilot as an integrated package, combining assisted planning, isolation support, logging, monitoring, and triaging. It does not isolate the contributions of individual components or distinguish the benefits of additional automation from those of their interface presentation. Component-level comparisons are needed to establish these mechanisms. Future work should deploy \pp in the wild with larger, more diverse groups and repeated use to examine how familiarity shapes practice and how coordination extends to code review, reconciliation, and integration.

Finally, our studies capture rapidly evolving practices of parallel AI coding: the formative study took place in mid-June 2026 and the controlled study in mid-August 2026. Participants developed strategies through social learning or repeated usage. While our findings identify coordination and monitoring challenges in current parallel AI coding, their specific manifestations may change as models, tools, and organizational norms evolve. Future work should examine how these challenges and developers’ strategies develop over time.

\section{Conclusion}
Parallel AI coding turns developers' waiting time into active
supervision of concurrent agents. Through formative interviews
($N=14$), we identified PILOT: five supervisory practices for
Planning, Isolating, Logging, Observing, and Triaging parallel
sessions. \pp{} turned these practices into interface support, and in a controlled study ($N=16$), participants completed more tickets, ran more agents at once, and spent less effort tracking them.
These results offer three takeaways. 
For tool designers, PILOT is a scaffold: each pillar names where parallel supervision breaks down and what interface support can carry that load. For builders of coding assistants, supervision should be a first-class layer around agents rather than an afterthought in a chat window, spanning dependency-aware plans before launch, persistent memory during execution, and triage cues that separate an agent's reported finish from the developer's acceptance. More broadly, PILOT frames a general problem of distributing limited attention across concurrent work, one likely to recur wherever people delegate in parallel.
Our null result on user control marks the next frontier. Participants knew when agents needed them but not always how to steer them, and some felt less familiar with the code once the tool took over the tracking they used to do themselves. Future tools should treat the abstraction level as a design choice rather than a fixed benefit of automation, and design for re-engagement: a short path from status back to the code, and the evidence needed to judge and redirect agents. Supervision support should not only reduce overhead but also keep developers ready to take control.

\bibliographystyle{ACM-Reference-Format}
\bibliography{3_bib}

@misc{openai_understanding_nodate,
	title = {Understanding and counting tokens},
	url = {https://help.openai.com/en/articles/4936856-understanding-and-counting-tokens},
	language = {en-US},
    year={2026},
	urldate = {2026-09-05},
	journal = {OpenAI Help Center},
	author = {OpenAI},
}

@inproceedings{overtrust_aicodingagents,
  author = {Yardim, Asli and Serafini, Raphael and Jost, Nadine and Ortloff, Anna-Marie and Gabriel Speckels, Joshua and Naiakshina, Alena},
title = {"The AI tool can’t make it any worse." Investigating Developers’ Security Behavior with AI Assistants in a Password Storage Study},
year = {2026},
isbn = {9798400722783},
publisher = {Association for Computing Machinery},
address = {New York, NY, USA},
url = {https://doi.org/10.1145/3772318.3791693},
doi = {10.1145/3772318.3791693},
booktitle = {Proceedings of the 2026 CHI Conference on Human Factors in Computing Systems},
articleno = {1216},
numpages = {33},
location = {
},
series = {CHI '26}
}

@misc{overtrust_notreading_AISabotage,
	title = {Coding with "{Enemy}": {Can} {Human} {Developers} {Detect} {AI} {Agent} {Sabotage}?},
	shorttitle = {Coding with "{Enemy}"},
	url = {http://arxiv.org/abs/2606.05647},
	doi = {10.48550/arXiv.2606.05647},
	urldate = {2026-09-05},
	publisher = {arXiv},
	author = {Ye, Jingheng and Zou, Huiqi and Yu, Simon and Shi, Weiyan},
	month = jun,
	year = {2026},
	note = {arXiv:2606.05647 [cs.AI]
version: 1},
}

@misc{overtrust_morecode_lessverification,
	title = {More code, less validation: {Risk} factors for over-reliance on {AI} coding tools among scientists},
	shorttitle = {More code, less validation},
	url = {http://arxiv.org/abs/2512.19644},
	doi = {10.48550/arXiv.2512.19644},
	urldate = {2026-09-05},
	publisher = {arXiv},
	author = {O'Brien, Gabrielle and Parker, Alexis and Eisty, Nasir and Carver, Jeffrey},
	month = dec,
	year = {2025},
	note = {arXiv:2512.19644 [cs.SE]
version: 1},
}

@misc{parallel_addcoordination14imporve,
	title = {When {Parallelism} {Pays} {Off}: {Cohesion}-{Aware} {Task} {Partitioning} for {Multi}-{Agent} {Coding}},
	shorttitle = {When {Parallelism} {Pays} {Off}},
	url = {http://arxiv.org/abs/2606.00953},
	doi = {10.48550/arXiv.2606.00953},
	urldate = {2026-09-05},
	publisher = {arXiv},
	author = {Yang, Xu and Nie, Lunyiu and Chandra, Ethan and Gannutin, Stanislav and Lin, Fangru and Chaudhuri, Swarat},
	month = may,
	year = {2026},
	note = {arXiv:2606.00953 [cs.LG]},
}

@inproceedings{log_develoeprNotRead,
	title = {"{I}'m {Not} {Reading} {All} of {That}": {Understanding} {Software} {Engineers}' {Level} of {Cognitive} {Engagement} with {Agentic} {Coding} {Assistants}},
	shorttitle = {"{I}'m {Not} {Reading} {All} of {That}"},
	url = {http://arxiv.org/abs/2603.14225},
	doi = {10.48550/arXiv.2603.14225},
	urldate = {2026-09-05},
	publisher = {arXiv},
	author = {Catalan, Carlos Rafael and Dizon, Lheane Marie and Monderin, Patricia Nicole and Kuang, Emily},
	month = mar,
	year = {2026},
	note = {arXiv:2603.14225 [cs.HC]},
	booktitle = {CHI 2026 Workshop on Tools for Thought},
}

@article{distributedcognition_old,
  title={Distributed cognition in an airline cockpit},
  author={Hutchins, Edwin and Klausen, Tove},
  journal={Cognition and communication at work},
  volume={1},
  pages={15--34},
  year={1996},
  publisher={Cambridge University Press, New York}
}

@article{distributedcognition_new,
author = {Hollan, James and Hutchins, Edwin and Kirsh, David},
title = {Distributed cognition: toward a new foundation for human-computer interaction research},
year = {2000},
issue_date = {June 2000},
publisher = {Association for Computing Machinery},
address = {New York, NY, USA},
volume = {7},
number = {2},
issn = {1073-0516},
url = {https://doi.org/10.1145/353485.353487},
doi = {10.1145/353485.353487},
journal = {ACM Trans. Comput.-Hum. Interact.},
month = jun,
pages = {174–196},
numpages = {23}
}

@misc{developer_log,
	title = {You should keep a developer’s journal - {Stack} {Overflow}},
	url = {https://stackoverflow.blog/2024/12/24/you-should-keep-a-developer-s-journal/},
	language = {en},
	urldate = {2026-09-08},
	author = {Pekarsky, Max},
	month = dec,
	year = {2024},
}

@misc{contextswitching_log,
	title = {Evolving with {AI}: {A} {Longitudinal} {Analysis} of {Developer} {Logs}},
	shorttitle = {Evolving with {AI}},
	url = {http://arxiv.org/abs/2601.10258},
	doi = {10.1145/3744916.3787811},
	urldate = {2026-09-08},
	author = {Sergeyuk, Agnia and Huang, Eric and Karaeva, Dariia and Serova, Anastasiia and Golubev, Yaroslav and Ahmed, Iftekhar},
	month = mar,
	year = {2026},
	note = {arXiv:2601.10258 [cs.SE]},
}

@misc{agent_orchestration3,
	title = {{AgentOrchestra}: {Orchestrating} {Multi}-{Agent} {Intelligence} with the {Tool}-{Environment}-{Agent}({TEA}) {Protocol}},
	shorttitle = {{AgentOrchestra}},
	url = {http://arxiv.org/abs/2506.12508},
	doi = {10.48550/arXiv.2506.12508},
	urldate = {2026-09-08},
	publisher = {arXiv},
	author = {Zhang, Wentao and Zeng, Liang and Xiao, Yuzhen and Li, Yongcong and Cui, Ce and Zhao, Yilei and Hu, Rui and Liu, Yang and Zhou, Yahui and An, Bo},
	month = may,
	year = {2026},
	note = {arXiv:2506.12508 [cs.AI]},
}

@misc{agent_orchestration2,
	title = {{AgentCoord}: {Visually} {Exploring} {Coordination} {Strategy} for {LLM}-based {Multi}-{Agent} {Collaboration}},
	shorttitle = {{AgentCoord}},
	url = {http://arxiv.org/abs/2404.11943},
	doi = {10.48550/arXiv.2404.11943},
	urldate = {2026-09-08},
	publisher = {arXiv},
	author = {Pan, Bo and Lu, Jiaying and Wang, Ke and Zheng, Li and Wen, Zhen and Feng, Yingchaojie and Zhu, Minfeng and Chen, Wei},
	month = apr,
	year = {2024},
	note = {arXiv:2404.11943 [cs.HC]},
}

@misc{agent_orchestration,
	title = {Multi-{Agent} {Collaboration} via {Evolving} {Orchestration}},
	url = {http://arxiv.org/abs/2505.19591},
	doi = {10.48550/arXiv.2505.19591},
	urldate = {2026-09-08},
	publisher = {arXiv},
	author = {Dang, Yufan and Qian, Chen and Luo, Xueheng and Fan, Jingru and Xie, Zihao and Shi, Ruijie and Chen, Weize and Yang, Cheng and Che, Xiaoyin and Tian, Ye and Xiong, Xuantang and Han, Lei and Liu, Zhiyuan and Sun, Maosong},
	month = oct,
	year = {2025},
	note = {arXiv:2505.19591 [cs.CL]},
}

@inproceedings{parallel_opportunty,
author = {Sch{\"o}mbs, Sarah and Zhang, Yan and Goncalves, Jorge and Johal, Wafa},
title = {From Conversation to Orchestration: HCI Challenges and Opportunities in Interactive Multi-Agentic Systems},
year = {2026},
isbn = {9798400721786},
publisher = {Association for Computing Machinery},
address = {New York, NY, USA},
url = {https://doi.org/10.1145/3765766.3765795},
doi = {10.1145/3765766.3765795},
booktitle = {Proceedings of the 13th International Conference on Human-Agent Interaction},
pages = {158–168},
numpages = {11},
location = {
},
series = {HAI '25}
}

@inproceedings{proactivity_mozannar2024show,
  title={When to show a suggestion? Integrating human feedback in AI-assisted programming},
  author={Mozannar, Hussein and Bansal, Gagan and Fourney, Adam and Horvitz, Eric},
  booktitle={Proceedings of the AAAI conference on artificial intelligence},
  volume={38},
  number={9},
  pages={10137--10144},
  year={2024}
}

@inproceedings{proactivity_valerie,
author = {Chen, Valerie and Zhu, Alan and Zhao, Sebastian and Mozannar, Hussein and Sontag, David and Talwalkar, Ameet},
title = {Need Help? Designing Proactive AI Assistants for Programming},
year = {2025},
isbn = {9798400713941},
publisher = {Association for Computing Machinery},
address = {New York, NY, USA},
url = {https://doi.org/10.1145/3706598.3714002},
doi = {10.1145/3706598.3714002},
booktitle = {Proceedings of the 2025 CHI Conference on Human Factors in Computing Systems},
articleno = {881},
numpages = {18},
location = {
},
series = {CHI '25}
}

@inproceedings{feedquac,
author = {Long, Tao and Wannamaker, Kendra and Vermeulen, Jo and Fitzmaurice, George and Matejka, Justin},
title = {FeedQUAC: Quick Unobtrusive AI-Generated Commentary},
year = {2026},
isbn = {9798400721786},
publisher = {Association for Computing Machinery},
address = {New York, NY, USA},
url = {https://doi.org/10.1145/3765766.3765861},
doi = {10.1145/3765766.3765861},
booktitle = {Proceedings of the 13th International Conference on Human-Agent Interaction},
pages = {452–455},
numpages = {4},
location = {
},
series = {HAI '25}
}

@inproceedings{taolongitudinal,
author = {Long, Tao and Gero, Katy Ilonka and Chilton, Lydia B},
title = {Not Just Novelty: A Longitudinal Study on Utility and Customization of an AI Workflow},
year = {2024},
isbn = {9798400705830},
publisher = {Association for Computing Machinery},
address = {New York, NY, USA},
url = {https://doi.org/10.1145/3643834.3661587},
doi = {10.1145/3643834.3661587},
booktitle = {Proceedings of the 2024 ACM Designing Interactive Systems Conference},
pages = {782–803},
numpages = {22},
location = {Copenhagen, Denmark},
series = {DIS '24}
}

@inproceedings{distributedcognition_doubleagents,
author = {Long, Tao and Zhang, Xuanming and Wang, Sitong and Yu, Zhou and Chilton, Lydia B},
title = {DoubleAgents: Human-Agent Alignment in a Socially Embedded Workflow},
year = {2026},
isbn = {9798400728945},
publisher = {Association for Computing Machinery},
address = {New York, NY, USA},
url = {https://doi.org/10.1145/3834580.3838737},
doi = {10.1145/3834580.3838737},
booktitle = {Proceedings of the 2026 ACM Conference on Human-AI Complementarity and Alignment},
pages = {23–44},
numpages = {22},
location = {
},
series = {HCOMP '26}
}

@misc{agent_comm_challenge,
	title = {Challenges in {Human}-{Agent} {Communication}},
	url = {http://arxiv.org/abs/2412.10380},
	doi = {10.48550/arXiv.2412.10380},
	urldate = {2026-09-08},
	publisher = {arXiv},
	author = {Bansal, Gagan and Vaughan, Jennifer Wortman and Amershi, Saleema and Horvitz, Eric and Fourney, Adam and Mozannar, Hussein and Dibia, Victor and Weld, Daniel S.},
	month = nov,
	year = {2024},
	note = {arXiv:2412.10380 [cs.HC]},
}

@inproceedings{authenticity_Med,
  title={" The human body is a black box" supporting clinical decision-making with deep learning},
  author={Sendak, Mark and Elish, Madeleine Clare and Gao, Michael and Futoma, Joseph and Ratliff, William and Nichols, Marshall and Bedoya, Armando and Balu, Suresh and O'Brien, Cara},
  booktitle={Proceedings of the 2020 conference on fairness, accountability, and transparency},
  pages={99--109},
  year={2020}
}

@inproceedings{authenticity_cocreativeKaty,
author = {Gero, Katy Ilonka and Long, Tao and Schnitzler, Carly and Dhillon, Paramveer},
title = {From Planning to Revision: How AI Writing Support at Different Stages Alters Ownership},
year = {2026},
isbn = {9798400725630},
publisher = {Association for Computing Machinery},
address = {New York, NY, USA},
url = {https://doi.org/10.1145/3800645.3813003},
doi = {10.1145/3800645.3813003},
booktitle = {Proceedings of the 2026 Designing Interactive Systems Conference},
pages = {2903–2934},
numpages = {32},
location = {
},
series = {DIS '26}
}

@article{authenticity_cocreativeAngel,
author = {Hwang, Angel Hsing-Chi and Liao, Q. Vera and Blodgett, Su Lin and Olteanu, Alexandra and Trischler, Adam},
title = { 'It was 80\% me, 20\% AI': Seeking Authenticity in Co-Writing with Large Language Models},
year = {2025},
issue_date = {May 2025},
publisher = {Association for Computing Machinery},
address = {New York, NY, USA},
volume = {9},
number = {2},
url = {https://doi.org/10.1145/3711020},
doi = {10.1145/3711020},
journal = {Proc. ACM Hum.-Comput. Interact.},
month = may,
articleno = {CSCW122},
numpages = {41}
}

@article{situationalawareness,
	title = {A {Situation} {Awareness} {Perspective} on {Human}-{AI} {Interaction}: {Tensions} and {Opportunities}},
	volume = {39},
	issn = {1044-7318},
	shorttitle = {A {Situation} {Awareness} {Perspective} on {Human}-{AI} {Interaction}},
	url = {https://doi.org/10.1080/10447318.2022.2093863},
	doi = {10.1080/10447318.2022.2093863},
	number = {9},
	urldate = {2026-09-08},
	journal = {International Journal of Human–Computer Interaction},
	publisher = {Taylor \& Francis},
	author = {Jiang, Jinglu and Karran, Alexander J. and Coursaris, Constantinos K. and Léger, Pierre-Majorique and Beringer, Joerg},
	month = may,
	year = {2023},
	pages = {1789--1806},
}

@misc{challenges_concurrent_dependencies,
	title = {Effective {Strategies} for {Asynchronous} {Software} {Engineering} {Agents}},
	url = {http://arxiv.org/abs/2603.21489},
	doi = {10.48550/arXiv.2603.21489},
	urldate = {2026-09-09},
	publisher = {arXiv},
	author = {Geng, Jiayi and Neubig, Graham},
	month = jul,
	year = {2026},
	note = {arXiv:2603.21489 [cs.CL]},
}

@InProceedings{motivation_actionupcodingcapability,
  title = 	 {Executable Code Actions Elicit Better {LLM} Agents},
  author =       {Wang, Xingyao and Chen, Yangyi and Yuan, Lifan and Zhang, Yizhe and Li, Yunzhu and Peng, Hao and Ji, Heng},
  booktitle = 	 {Proceedings of the 41st International Conference on Machine Learning},
  pages = 	 {50208--50232},
  year = 	 {2024},
  editor = 	 {Salakhutdinov, Ruslan and Kolter, Zico and Heller, Katherine and Weller, Adrian and Oliver, Nuria and Scarlett, Jonathan and Berkenkamp, Felix},
  volume = 	 {235},
  series = 	 {Proceedings of Machine Learning Research},
  month = 	 {21--27 Jul},
  publisher =    {PMLR},
  url = 	 {https://proceedings.mlr.press/v235/wang24h.html}
}

@misc{parallel_liu,
	title = {Agentic {Coding} in the {Wild}: {Characterizing} {GitHub} {Copilot} {Traces} at {Production} {Scale}},
	shorttitle = {Agentic {Coding} in the {Wild}},
	url = {http://arxiv.org/abs/2608.00101},
	doi = {10.48550/arXiv.2608.00101},
	urldate = {2026-09-09},
	publisher = {arXiv},
	author = {Liu, Banruo and Qiu, Haoran and Goiri, Íñigo and Fonseca, Rodrigo and Bianchini, Ricardo and Choukse, Esha},
	month = jul,
	year = {2026},
	note = {arXiv:2608.00101 [cs.AI]},
}

@misc{parallel_haijun,
	title = {Professional {Software} {Developers} {Don}'t {Vibe}, {They} {Control}: {AI} {Agent} {Use} for {Coding} in 2025},
	shorttitle = {Professional {Software} {Developers} {Don}'t {Vibe}, {They} {Control}},
	url = {http://arxiv.org/abs/2512.14012},
	doi = {10.48550/arXiv.2512.14012},
	urldate = {2026-09-09},
	publisher = {arXiv},
	author = {Huang, Ruanqianqian and Reyna, Avery and Lerner, Sorin and Xia, Haijun and Hempel, Brian},
	month = aug,
	year = {2026},
	note = {arXiv:2512.14012 [cs.SE]},
}

@inproceedings{autonomous_sweagent,
author = {Yang, John and Jimenez, Carlos E. and Wettig, Alexander and Lieret, Kilian and Yao, Shunyu and Narasimhan, Karthik and Press, Ofir},
title = {SWE-agent: agent-computer interfaces enable automated software engineering},
year = {2024},
isbn = {9798331314385},
publisher = {Curran Associates Inc.},
address = {Red Hook, NY, USA},
booktitle = {Proceedings of the 38th International Conference on Neural Information Processing Systems},
articleno = {1601},
numpages = {125},
location = {Vancouver, BC, Canada},
series = {NIPS '24}
}

@misc{concurrent_openaireport,
	title = {Research acceleration: {The} view inside {OpenAI}},
	shorttitle = {Research acceleration},
	url = {https://openai.com/index/research-acceleration-view-inside-openai/},
	language = {en-US},
	urldate = {2026-09-09},
	journal = {OpenAI},
	author = {OpenAI},
	month = sep,
	year = {2026},
}

@inproceedings{visibility_challeneg_facct,
	address = {New York, NY, USA},
	series = {{FAccT} '24},
	title = {Visibility into {AI} {Agents}},
	isbn = {979-8-4007-0450-5},
	url = {https://dl.acm.org/doi/10.1145/3630106.3658948},
	doi = {10.1145/3630106.3658948},
	urldate = {2026-09-08},
	booktitle = {Proceedings of the 2024 {ACM} {Conference} on {Fairness}, {Accountability}, and {Transparency}},
	publisher = {Association for Computing Machinery},
	author = {Chan, Alan and Ezell, Carson and Kaufmann, Max and Wei, Kevin and Hammond, Lewis and Bradley, Herbie and Bluemke, Emma and Rajkumar, Nitarshan and Krueger, David and Kolt, Noam and Heim, Lennart and Anderljung, Markus},
	month = jun,
	year = {2024},
	pages = {958--973},
}

@misc{visibility_example2,
	title = {{AgentLens}: {Visual} {Analysis} for {Agent} {Behaviors} in {LLM}-based {Autonomous} {Systems}},
	shorttitle = {{AgentLens}},
	url = {http://arxiv.org/abs/2402.08995},
	doi = {10.48550/arXiv.2402.08995},
	urldate = {2026-09-08},
	publisher = {arXiv},
	author = {Lu, Jiaying and Pan, Bo and Chen, Jieyi and Feng, Yingchaojie and Hu, Jingyuan and Peng, Yuchen and Chen, Wei},
	month = feb,
	year = {2024},
	note = {arXiv:2402.08995 [cs.HC]
version: 1},
}

@misc{visibility_example,
	title = {Agentic {Visualization}: {Extracting} {Agent}-based {Design} {Patterns} from {Visualization} {Systems}},
	shorttitle = {Agentic {Visualization}},
	url = {http://arxiv.org/abs/2505.19101},
	doi = {10.1109/MCG.2025.3607741},
	urldate = {2026-09-08},
	author = {Dhanoa, Vaishali and Wolter, Anton and León, Gabriela Molina and Schulz, Hans-Jörg and Elmqvist, Niklas},
	month = sep,
	year = {2025},
	note = {arXiv:2505.19101 [cs.HC]},
}

@misc{parallel_stoa,
	title = {{AI} {Factory}: {How} {One} {Developer} {Ships} 72 {Story} {Points}/{Day} {\textbar} {STOA} {Docs}},
	shorttitle = {{AI} {Factory}},
	url = {https://docs.gostoa.dev/blog/how-we-built-ai-factory-ships-72-points-per-day},
	language = {en-US},
	urldate = {2026-09-05},
	journal = {STOA Documentation},
	author = {STOA},
	month = feb,
	year = {2026},
}

@misc{parallel_cursor,
	title = {Faire doubles {PR} throughput with {Cursor} {Cloud} {Agents} · {Cursor}},
	url = {https://cursor.com/blog/faire},
	language = {en-US},
	urldate = {2026-09-05},
	journal = {Cursor},
	author = {Cursor},
	month = may,
	year = {2026},
}

@misc{parallel_onedevAI_do4times,
	title = {One {Developer} {Is} {All} {You} {Need}: {A} {Case} {Study} of an {AI}-{Augmented} {One}-{Person} {Squad} in a {Brownfield} {Enterprise}},
	shorttitle = {One {Developer} {Is} {All} {You} {Need}},
	url = {http://arxiv.org/abs/2605.18461},
	doi = {10.48550/arXiv.2605.18461},
	urldate = {2026-09-09},
	publisher = {arXiv},
	author = {Boas, Marcelo Vilas and Pinto, Gustavo and Monteiro, Edward Roberto and Carida, Vinicius Fernandes and Ribeiro, Danilo},
	year = {2026},
	note = {arXiv:2605.18461 [cs.SE]},
}

@inproceedings{parallel_github,
	address = {Austin Texas},
	title = {The sky is not the limit: multitasking across {GitHub} projects},
	isbn = {978-1-4503-3900-1},
	shorttitle = {The sky is not the limit},
	url = {https://dl.acm.org/doi/10.1145/2884781.2884875},
	doi = {10.1145/2884781.2884875},
	language = {en},
	urldate = {2026-09-05},
	booktitle = {Proceedings of the 38th {International} {Conference} on {Software} {Engineering}},
	publisher = {ACM},
	author = {Vasilescu, Bogdan and Blincoe, Kelly and Xuan, Qi and Casalnuovo, Casey and Damian, Daniela and Devanbu, Premkumar and Filkov, Vladimir},
	month = may,
	year = {2016},
	pages = {994--1005},
}

@article{readingspeed_238,
	title = {How many words do we read per minute? {A} review and meta-analysis of reading rate},
	volume = {109},
	issn = {0749-596X},
	url = {https://www.sciencedirect.com/science/article/pii/S0749596X19300786},
	doi = {https://doi.org/10.1016/j.jml.2019.104047},
	journal = {Journal of Memory and Language},
	author = {Brysbaert, Marc},
	year = {2019},
	pages = {104047},
}

@inproceedings{affinityDiagramming,
  title={Using affinity diagrams to evaluate interactive prototypes},
  author={Lucero, Andr{\'e}s},
  booktitle={IFIP conference on human-computer interaction},
  pages={231--248},
  year={2015},
  organization={Springer}
}

@article{thematicAnalysis,
  title={Demonstrating rigor using thematic analysis: A hybrid approach of inductive and deductive coding and theme development},
  author={Fereday, Jennifer and Muir-Cochrane, Eimear},
  journal={International journal of qualitative methods},
  volume={5},
  number={1},
  pages={80--92},
  year={2006},
  publisher={SAGE Publications Sage CA: Los Angeles, CA}
}

@article{benjamini_Hochberg,
  title={Quick and easy implementation of the Benjamini-Hochberg procedure for controlling the false positive rate in multiple comparisons},
  author={Thissen, David and Steinberg, Lynne and Kuang, Daniel},
  journal={Journal of educational and behavioral statistics},
  volume={27},
  number={1},
  pages={77--83},
  year={2002},
  publisher={Sage Publications Sage CA: Los Angeles, CA}
}

@incollection{theory_coordination_monitoring2,
	title = {Supervisory {Control}},
	isbn = {978-0-470-04820-7},
	url = {https://onlinelibrary.wiley.com/doi/abs/10.1002/0470048204.ch38},
	doi = {10.1002/0470048204.ch38},
	language = {en},
	urldate = {2026-09-09},
	booktitle = {Handbook of {Human} {Factors} and {Ergonomics}},
	publisher = {John Wiley \& Sons, Ltd},
	author = {Sheridan, Thomas B.},
	year = {2006},
	doi = {10.1002/0470048204.ch38},
	note = {Section: 38},
	pages = {1025--1052},
}

@inproceedings{socialdynamics,
author = {Gero, Katy Ilonka and Long, Tao and Chilton, Lydia B},
title = {Social Dynamics of AI Support in Creative Writing},
year = {2023},
isbn = {9781450394215},
publisher = {Association for Computing Machinery},
address = {New York, NY, USA},
url = {https://doi.org/10.1145/3544548.3580782},
doi = {10.1145/3544548.3580782},
booktitle = {Proceedings of the 2023 CHI Conference on Human Factors in Computing Systems},
articleno = {245},
numpages = {15},
location = {Hamburg, Germany},
series = {CHI '23}
}

@article{theory_coordination_monitoring,
	title = {Considerations in {Modeling} the {Human} {Supervisory} {Controller}},
	volume = {8},
	issn = {1474-6670},
	url = {https://www.sciencedirect.com/science/article/pii/S1474667017675554},
	doi = {https://doi.org/10.1016/S1474-6670(17)67555-4},
	number = {1, Part 3},
	journal = {IFAC Proceedings Volumes},
	author = {Sheridan, Thomas B.},
	year = {1975},
	pages = {223--228},
}

@article{oppotunty_se_changinglandscape,
author = {Hassan, Ahmed E. and Oliva, Gustavo A. and Lin, Dayi and Chen, Boyuan and Jiang, Zhen Ming (Jack)},
title = {Towards AI-Native Software Engineering (SE 3.0): A Vision and a Challenge Roadmap},
year = {2026},
issue_date = {September 2026},
publisher = {Association for Computing Machinery},
address = {New York, NY, USA},
volume = {35},
number = {9},
issn = {1049-331X},
url = {https://doi.org/10.1145/3807901},
doi = {10.1145/3807901},
journal = {ACM Trans. Softw. Eng. Methodol.},
month = aug,
articleno = {268},
numpages = {25}
}

@inproceedings{log_developeraddAgentMd,
	title = {On the {Impact} of {AGENTS}.md {Files} on the {Efficiency} of {AI} {Coding} {Agents}},
	url = {http://arxiv.org/abs/2601.20404},
	doi = {10.48550/arXiv.2601.20404},
	urldate = {2026-09-09},
	publisher = {arXiv},
	author = {Lulla, Jai Lal and Mohsenimofidi, Seyedmoein and Galster, Matthias and Zhang, Jie M. and Baltes, Sebastian and Treude, Christoph},
	month = mar,
	year = {2026},
	note = {arXiv:2601.20404 [cs.SE]},
}

@misc{log_tracelab_cantread,
	title = {{TraceLab}: {Characterizing} {Coding} {Agent} {Workloads} for {LLM} {Serving}},
	shorttitle = {{TraceLab}},
	url = {http://arxiv.org/abs/2606.30560},
	doi = {10.48550/arXiv.2606.30560},
	urldate = {2026-09-05},
	publisher = {arXiv},
	author = {Zhu, Kan and Jacob, Mathew and Ma, Chenxi and Pan, Yi and Wang, Stephanie and Krishnamurthy, Arvind and Kasikci, Baris},
	month = jun,
	year = {2026},
	note = {arXiv:2606.30560 [cs.LG]},
}

@misc{log_faster5xhumanreadability,
	title = {Your {Agent} {Writes} {Faster} {Than} {You} {Can} {Read}},
	url = {https://blakecrosley.com/blog/cognitive-debt-agents},
	language = {en},
	urldate = {2026-09-05},
	journal = {Blake Crosley},
	author = {Crosley, Blake},
	month = feb,
	year = {2026},
	note = {Section: AI \& Technology},
}

@online{claudeoutputMuch,
 author = {Zoe Hitzig and Maxim Massenkoff and Eva Lyubich and Shaoyi Zhang and Ryan Heller and Peter McCrory},
 title = {Agentic coding and persistent returns to expertise},
 date = {2026-06-16},
 year = {2026},
 url = {https://www.anthropic.com/research/claude-code-expertise},
}

@inproceedings{oversight_workshop,
author = {Khadar, Malik and Cecil, Julia and Van Der Neut, Leon and Banovic, Nikola and Baum, Kevin and Chancellor, Stevie and Costanza, Enrico and Eslami, Motahhare and Feit, Anna Maria and Gaube, Susanne and Gadiraju, Ujwal and Kaur, Harmanpreet},
title = {AI CHAOS! 2nd Workshop on the Challenges for Human Oversight of AI Systems},
year = {2026},
isbn = {9798400722813},
publisher = {Association for Computing Machinery},
address = {New York, NY, USA},
url = {https://doi.org/10.1145/3772363.3778736},
doi = {10.1145/3772363.3778736},
booktitle = {Proceedings of the Extended Abstracts of the 2026 CHI Conference on Human Factors in Computing Systems},
articleno = {919},
numpages = {6},
location = {
},
series = {CHI EA '26}
}

@misc{oversight_log_hard_to_read_Madeleine,
	title = {Overseeing {Agents} {Without} {Constant} {Oversight}: {Challenges} and {Opportunities}},
	copyright = {arXiv.org perpetual, non-exclusive license},
	shorttitle = {Overseeing {Agents} {Without} {Constant} {Oversight}},
	url = {https://arxiv.org/abs/2602.16844},
	doi = {10.48550/ARXIV.2602.16844},
	language = {en},
	urldate = {2026-09-04},
	publisher = {arXiv},
	author = {Grunde-McLaughlin, Madeleine and Mozannar, Hussein and Murad, Maya and Chen, Jingya and Amershi, Saleema and Fourney, Adam},
	year = {2026},
	note = {Version Number: 1},
}

@inproceedings{loghardtoreadoversight_produceoverwhelminginformation_paralleliz_ability_amplifies_the_oversight_burden,
	title = {Managing {Multi}-{Agent} {Research} {Systems}: {A} {Dashboard} for {Human} {Oversight} of {Coordinating} {AI} {Agents}},
	booktitle = {{Proceedings} of {Human}-centered {Evaluation} and {Auditing} of {Language} {Models} ({HEAL}@{CHI}’26). {ACM}, {New} {York}, {NY}, {USA}},
	language = {en},
	author = {Kitano, Brian and Carlson, Evan and Russett, Bryan and Kesling, Alex},
	year = {2026},
}

@inproceedings{viz_hardtoreadfromlongconversationwhendebuggingmultiagent,
author = {Epperson, Will and Bansal, Gagan and Dibia, Victor C and Fourney, Adam and Gerrits, Jack and Zhu, Erkang (Eric) and Amershi, Saleema},
title = {Interactive Debugging and Steering of Multi-Agent AI Systems},
year = {2025},
isbn = {9798400713941},
publisher = {Association for Computing Machinery},
address = {New York, NY, USA},
url = {https://doi.org/10.1145/3706598.3713581},
doi = {10.1145/3706598.3713581},
booktitle = {Proceedings of the 2025 CHI Conference on Human Factors in Computing Systems},
articleno = {156},
numpages = {15},
location = {
},
series = {CHI '25}
}

@misc{viz_traceishardtounderstandCUA,
      title={"What Did It Actually Do?": Understanding Risk Awareness and Traceability for Computer-Use Agents}, 
      author={Zifan Peng and Mingchen Li},
      year={2026},
      eprint={2603.28551},
      archivePrefix={arXiv},
      primaryClass={cs.CR},
      url={https://arxiv.org/abs/2603.28551}, 
}

@article{loghardtoread,
  title={Xai for coding agent failures: Transforming raw execution traces into actionable insights},
  author={Joshi, Arun},
  journal={arXiv preprint arXiv:2603.05941},
  year={2026}
}

\appendix
\section*{APPENDICES}


\section{\pp{} System Implementation Details}
\label{app:system-implementation}

\subsection{Planning Interface  Implementation Details}
\label{app:planning-implementation}

\paragraph{Task dependency analysis and plan generation.}
The planning backend processes the user's goal, ticket descriptions,
specification files, and repository context through chained calls to
\texttt{\small gpt-5.6-sol} with medium reasoning effort, balancing dependency-analysis quality with latency. These calls analyze task
breakdowns, dependencies, and potential overlapping edits, distinguishing
two constraint types: \emph{output dependencies}, which require a
particular execution order, and \emph{shared-file conflicts}, which
require non-overlapping work, before proposing agent assignments
and execution rounds. The results populate a dependency graph, an
editable drag-and-drop Gantt chart, and natural-language execution
instructions. Execution rounds, presented as waves, indicate task
order. A task can proceed
once its own prerequisites have been reviewed and confirmed, without
waiting for the unrelated work in other windows.

\paragraph{Plan revision and session setup.}
Informed by our formative study, the default plan uses three
parallel agents. Users can revise the plan directly or through
the planning assistant, which uses the same model and reasoning
effort as the planning backend. Model-generated dependency and
assignment recommendations remain open to user review and revision.
Clicking \button{Launch Agents $\rightarrow$} opens the setup page
and generates per-agent launch commands, jumpstart prompts, and
steering files from the supplied specifications, including
\texttt{\small AGENTS.md}, \texttt{\small CLAUDE.md}, and Copilot
instructions. Each agent works in an isolated repository directory.
The launch workflow associates sessions with their assigned tasks
and window identifiers, connecting the approved plan to subsequent
logging and monitoring.

\subsection{Run-Logger  Implementation Details}
\label{app:logger-implementation}

\paragraph{Capture and local persistence.}
The run-logger maintains three locally stored layers:
raw, full-length event logs (\texttt{\small.json}); mid-length, cross-session Markdown memory (\texttt{\small.md}); and short UI-card summaries of key events (\texttt{\small.md}). Raw events and conversation records preserve
structured agent activity, captured prompts and replies, and
timestamps. The \texttt{\small.json} records are persisted using atomic file replacement.
For Copilot, a background capture process polls the agent's SQLite
session store in read-only mode every two seconds. Conversation
turns are additionally retained in append-only Markdown and JSONL
archives that survive session resets. Logging and capture run in
background Node.js processes, using the filesystem for persistence
and read-only SQLite access for Copilot conversations.

\paragraph{Session memory and tracking summaries.}
When new conversation turns are detected, the system triggers
memory and summary generation using \texttt{\small gpt-5-mini}.
Mid-length session memory records context, decisions, progress,
and key events, including intervention snippets and complete
messages. Short tracking summaries are generated from the full
log and provide one-line progress updates and key events for
dashboard cards.
Session memory is accessible through the agent detail view and
can be reused as context for subsequent sessions or future
\pp{} runs. Raw records remain available locally for detailed
inspection, while the dashboard refreshes its view of the local
records approximately every 2.5 seconds.
 This design separates context preservation from the agent’s primary workflow: users do not need to
interrupt an agent or ask it to produce a separate recap. We see that this lightweight approach can support other CLI
agents through compatible logging integrations.

\subsection{Ambient Dashboard  Implementation Details}
\label{app:dashboard-implementation}


\paragraph{State updates and plan visualization.}
The dashboard maps the approved plan into task cards organized
by agent window and execution stage: \textit{Done},
\textit{Running Now}, and \textit{Next}. It polls the run-logger's
outputs and updates the interface through JavaScript approximately
every 2.5 seconds, combining structured status and file-change
events with generated tracking summaries. Planning-stage Spotlight
selections remain visible to help developers prioritize sessions
when coordinating subsequent work.

\paragraph{Intervention detection and review.}
A triage module uses \texttt{\small gpt-5.6-sol} with medium
reasoning effort to identify four potential intervention types from raw logs:
prolonged inactivity, terminal errors, requests for user input,
and completion messages. The first three trigger a
\texttt{\small needs\_attention} status, a glowing and blinking
card, and a one-line explanation.
Completion messages follow a separate path: the dashboard displays
the agent's reported completion and prompts the developer to verify
the result before moving the task to \textit{Done}. This separates
an agent's reported finish from the developer's acceptance of its
work; the completion cue is not itself a correctness check.
Direct links to terminal sessions and Markdown history support
inspection and intervention. Developers answer questions and
review results in the underlying CLI sessions rather than through
a replacement agent interface.

\section{User Study Method Details}

\subsection{Task Design Details}
\label{app:task-details}
We designed two software projects to approximate realistic development experiences, each presented through a mock GitHub issue page containing six coding tickets with explicit requirements and completion criteria. Sparkmatch was an almost empty dating application. Its tasks covered compatibility-survey collection, a compatibility-matching algorithm, a ranked recommendation feed, keyboard-driven like/pass interactions and mutual-match detection, a match overlay and matches list, and profile viewing and editing. Kart was a partially implemented social shopping application with prebuilt authentication and friend-purchase lookup functionality. Its tasks covered profile display, login-history recording, login-gated checkout, catalog browsing and product recommendation with friend-purchase indicators like ``N of your friends bought this,'' shopping-cart operations, and password changes.

We aligned the projects on task counts and key coordination features: independently implementable tasks, dependency relationships requiring cross-task integration, one deliberately shared-file task pair, and one task requiring an explicit product or policy decision. The shared-file pairs were swipe interactions and match display in Sparkmatch, and checkout and cart display in Kart; the dependency chains connected preference management to matching in Sparkmatch and browsing history to recommendations in Kart. The decision-oriented tasks concerned matching policy and catalog presentation, respectively. These structures created opportunities to exercise separability, coordinate dependent work, and resolve design choices with coding agents in parallel, rather than through full AI automation. To focus on coordination and monitoring during implementation, participants in both conditions were asked to address the tickets without being required to manage pull requests, review code, or merge changes.  

\subsection{Procedure Details}
\label{app:method-details}

\begin{figure}[!h]
    \centering
      \vspace{-10px}
    \includegraphics[width=\linewidth]{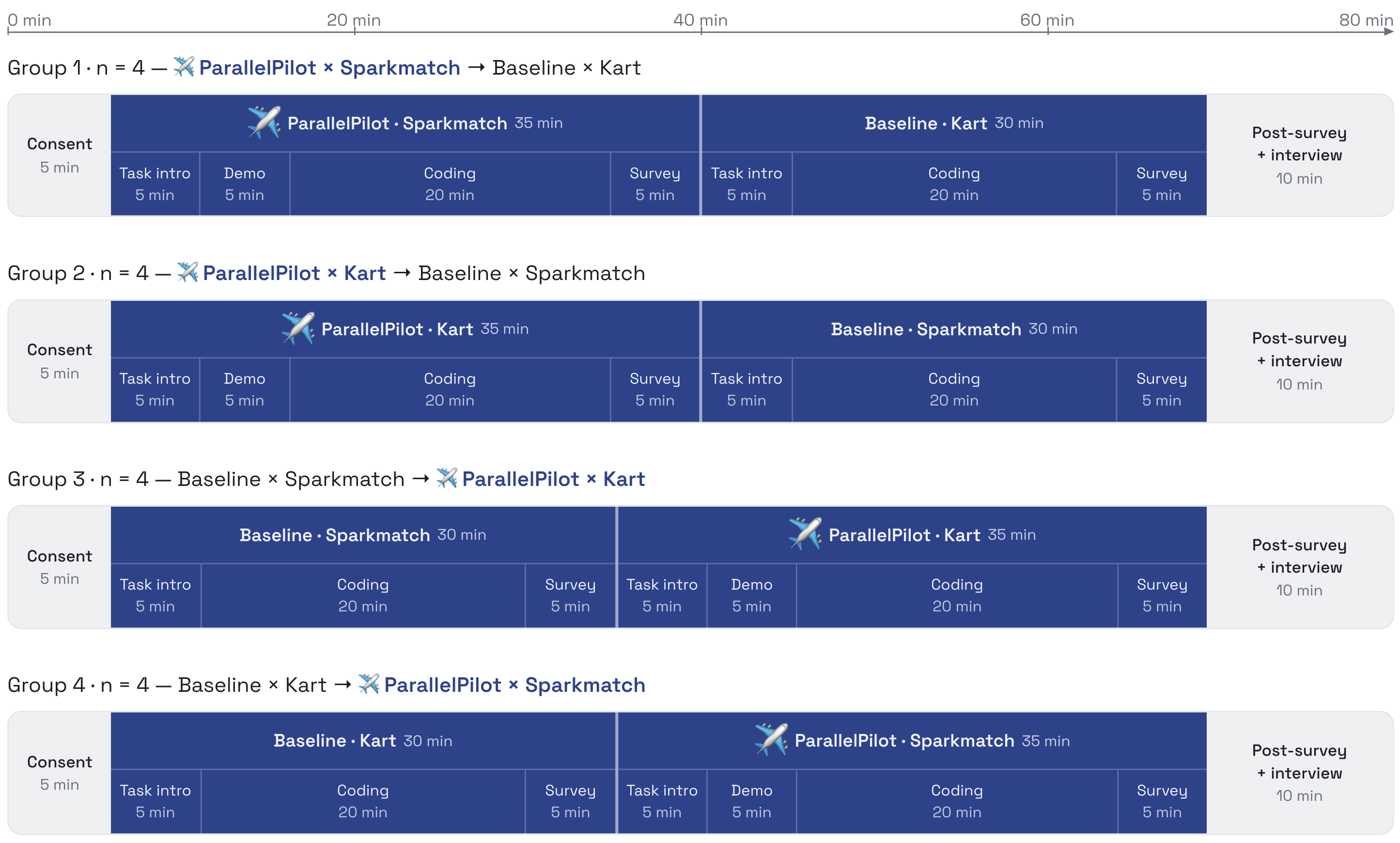}
    \caption{Session procedure for the within-subjects study ($N=16$, four groups of $n=4$). Each 80-minute session opens with a 5-minute consent and setup briefing, then two counterbalanced coding blocks that cross tool condition (\pp{} vs.\ baseline without ParallelPilot) with project (Sparkmatch vs.\ Kart), and closes with a 10-minute post-study survey and interview.}
  \Description{A horizontal timeline with a 0 to 80 minute axis at the top and four rows below it, one per participant group. Each row shows a sequence of blocks sized to their duration in minutes: a short block for consent and setup, two adjacent blocks for the two coding conditions, and a final block for the post-study survey and interview. Each coding block is subdivided into a task briefing segment, an onboarding video segment shown only for the \pp condition, a coding segment, and a survey segment. A plane icon marks which of the two coding blocks in each row is the \pp condition. The four rows differ only in the order of the two conditions and which project, Sparkmatch or Kart, is paired with which condition, illustrating the study's counterbalancing scheme.}
  \label{fig:procedure}
  \vspace{-10px}
\end{figure}

\subsection{Apparatus Details}
\label{app:apparatus-details}
Participants used a researcher-provided laptop (Windows 11, 32GB RAM) connected to two 23-inch extended monitors, creating a three-screen workspace. They could freely arrange and move windows across screens as needed, simulating a realistic multi-monitor development environment. The stack included VS Code with GitHub Copilot (chat and CLI, default model \texttt{{\small gpt-5.6-sol}} medium reasoning, access date: mid-August 2026), two sandboxed repositories (Sparkmatch and Kart), and GitHub-style ticket lists with descriptions and epics, available in VS Code and as HTML. Each condition starts and concludes with a logging and cleaning command. In the \pp condition, the planning interface and the monitoring dashboard ran locally. 
Each session used a newly authenticated Copilot instance, isolated exclusively to that participant; no authentication state, chat history, or context memory persisted across sessions or participants, ensuring realistic behavior without carryover.

\section{User Study Result Details}
\subsection{Study Design Checks: Task Equivalence}
\label{app:study-checks-tasks}

Perceived difficulty ratings and supplementary equivalence checks on
survey composites both point to broad comparability between tasks and
sessions, though not perfect equivalence. 
In the post-study survey, 9
of 16 participants rated the two projects as similarly hard; the
remaining 7 split on which was harder (3 Sparkmatch, 4 Kart).

We additionally computed five multi-item composites from the
two post-condition surveys (one per task) and tested them for
equivalence, as shown in Table~\ref{tab:task-equivalence}. Each composite
averaged its constituent items within a survey, and we paired
each participant's Kart and Sparkmatch scores for that composite
across tool assignments. Under TOST (two one-sided tests) with a $\pm1$-point margin on
the 1--7 scale, four of the five composites met equivalence,
with 90\% CIs falling entirely within the margin; the fifth,
supervision-specific load, was inconclusive, with too wide a CI
to establish either equivalence or a reliable difference. 

\begin{table}[!h]
\centering
\footnotesize
\setlength{\tabcolsep}{4pt}
\caption{Paired task comparisons ($N=16$).
$\Delta=\text{Kart}-\text{Sparkmatch}$.
Confidence intervals shown are 95\%; TOST equivalence decisions
use 90\% intervals and a $\pm1$ margin.}
\label{tab:task-equivalence}
\begin{tabular}{@{}p{.42\linewidth}rrrlp{.14\linewidth}@{}}
\toprule
\textbf{Composite (item count)} &
\textbf{Kart} &
\textbf{Sparkmatch} &
\textbf{$\Delta$ [95\% CI]} &
\textbf{TOST} \\
\midrule
Supervision-specific load (3)
& 3.54 & 3.33 & $+0.21$ [$-0.94$, $1.35$]
& Inconclusive \\

Awareness, memory, and recall (3)
& 4.61 & 4.42 & $+0.19$ [$-0.72$, $1.10$]
& Equivalent \\

Planning, decomposition, and delegation (5)
& 4.54 & 4.52 & $+0.01$ [$-0.82$, $0.84$]
& Equivalent \\

Control, verification, and intervention (4)
& 3.76 & 4.31 & $-0.55$ [$-0.98$, $-0.12$]
& Equivalent \\

Usability and experience (5)
& 4.65 & 4.68 & $-0.03$ [$-1.07$, $1.02$]
& Equivalent \\
\bottomrule
\end{tabular}
\end{table}

\subsection{Study Design Checks: Session-Order}
\label{app:study-checks-order}

We also compared first- and second-session composites using
paired tests, as shown in Table~\ref{tab:session-order}. None of the five
comparisons was statistically significant (all $p>.05$);
observed standardized differences were small
($|d_z|\leq0.29$). We therefore detected no session-order
differences in these subjective measures, although the small
sample does not rule out learning or carryover effects.

\begin{table}[!h]
\centering
\footnotesize
\setlength{\tabcolsep}{5pt}
\caption{Paired session-order comparisons ($N=16$).
$\Delta=\text{Session 2}-\text{Session 1}$;
$d_z$ is the paired standardized mean difference.
All comparisons had $p>.05$.}
\label{tab:session-order}
\begin{tabular}{@{}p{.43\linewidth}rrrr@{}}
\toprule
\textbf{Composite (item count)} &
\textbf{Session 1} &
\textbf{Session 2} &
\textbf{$\Delta$} &
\textbf{$d_z$} \\
\midrule
Supervision-specific load (3)
& 3.60 & 3.27 & $-0.33$ & $-0.16$ \\

Awareness, memory, and recall (3)
& 4.39 & 4.65 & $+0.26$ & $0.16$ \\

Planning, decomposition, and delegation (5)
& 4.40 & 4.66 & $+0.26$ & $0.17$ \\

Control, verification, and intervention (4)
& 3.90 & 4.17 & $+0.27$ & $0.29$ \\

Usability and experience (5)
& 4.69 & 4.64 & $-0.05$ & $-0.03$ \\
\bottomrule
\end{tabular}
\vspace{-20px}
\end{table}

\subsection{Task-Phase Time Allocation}
\label{app:phase-time}

\begin{figure}[!h]
  \centering
  \includegraphics[width=\linewidth]{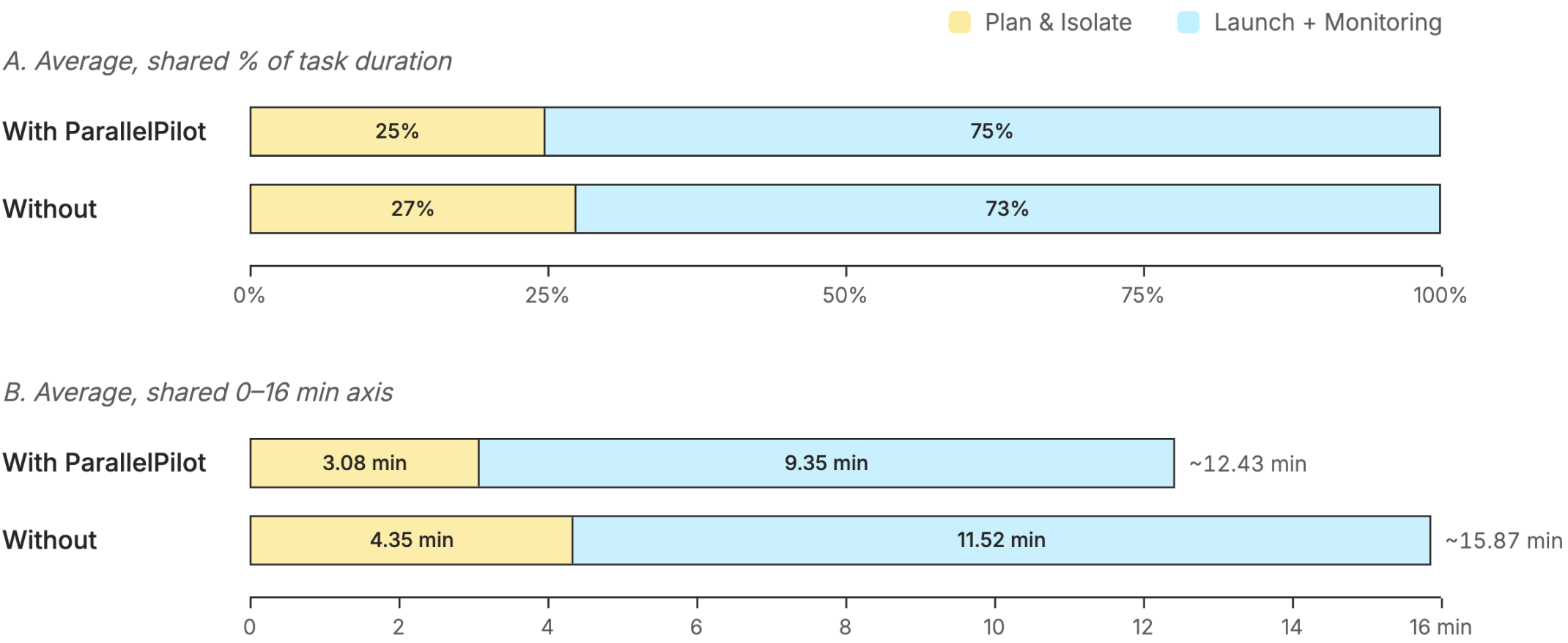}
  \caption{Task-phase time allocation for dual-completers with and without \pp. Panel~A gives the mean
  share of task duration spent on \emph{Plan \& Isolate} versus \emph{Launch +
  Monitoring}; Panel~B places the same means on a shared 0--16 min axis. }
  \label{fig:phase-time}
\end{figure}

Figure~\ref{fig:phase-time} shows that dual completers — the
participants who completed all 12 tasks in both conditions —
maintain a similar split of task-phase time while saving time
overall with \pp. Each of their tasks was coded into two
phases: \emph{Plan \& Isolate}, from task start to the first
launch, and \emph{Launch + Monitoring}, from that launch to
task completion.
Panel~A shows that the proportional split is
nearly unchanged across conditions: Plan \& Isolate accounts for
25\% of task time with \pp\ versus 27\% without, with Launch +
Monitoring making up the remainder. Panel~B shows that both
phases are shorter in absolute terms with \pp, by 1.27 and 2.17
minutes respectively, for a mean total of 12.43 minutes against
15.87 minutes. The tildes in Panel~B mark means rather than
observed task ends.

\subsection{Participant Interactions with ParallelPilot’s Three Components}
\label{app:user-study-setup-details}

\begin{figure*}[!t]
    \centering
    \includegraphics[width=.98\textwidth]{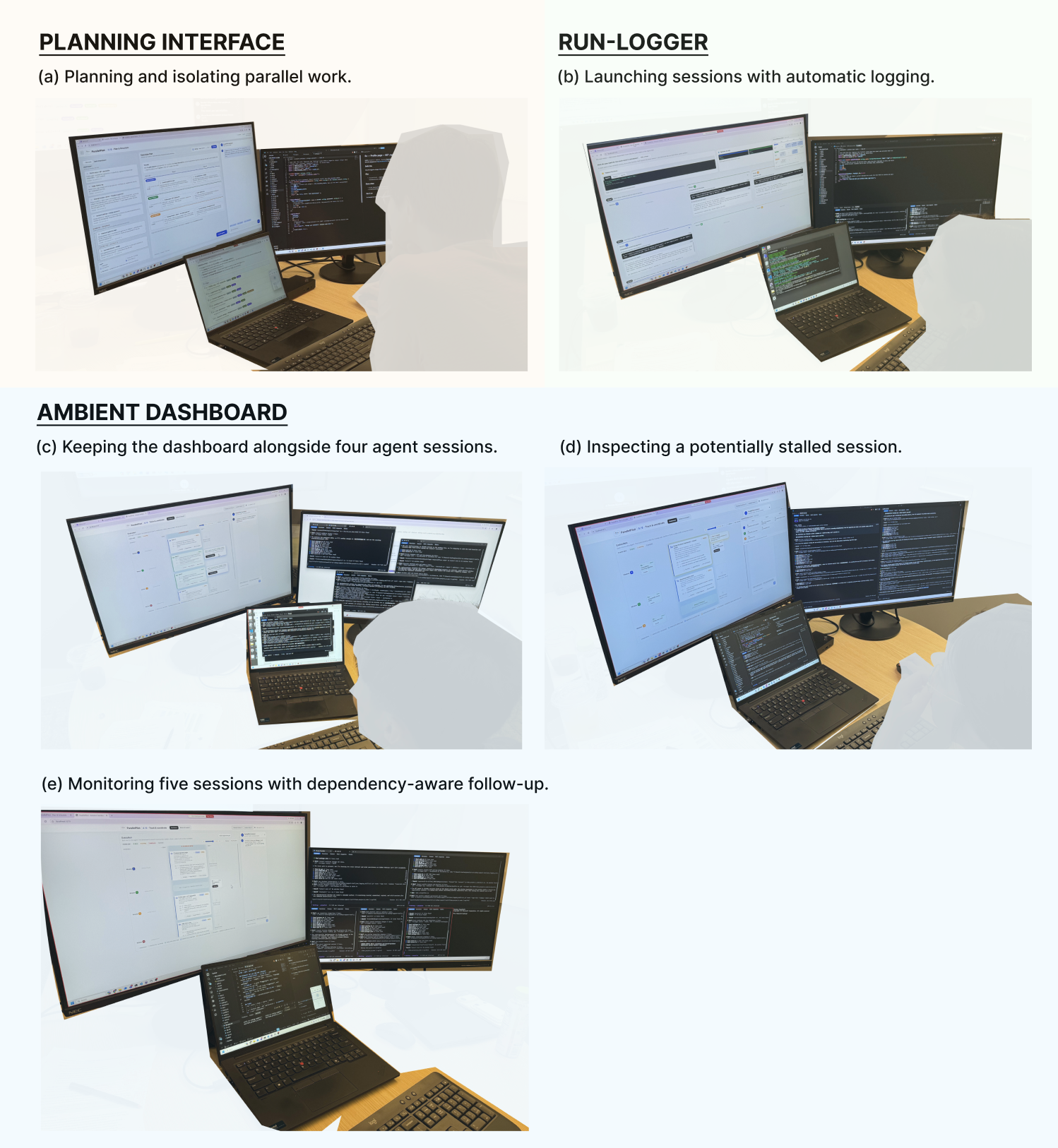}
    \caption{{Participants' use of ParallelPilot's three
    components alongside existing coding tools.}
    (a) \textbf{Planning interface:} reviewing the execution plan
    and dependencies on the left external monitor, with code
    on the right monitor and project tickets on the laptop.
    (b) \textbf{Run-logger:} following launch instructions on the
    left monitor to start sessions with automatic logging,
    alongside VS Code terminals on the right monitor and
    a terminal on the laptop.
(c--e) \textbf{Ambient dashboard:} the left monitor keeps overall progress visible while the user (c) receives two completion notifications for four sessions running concurrently across the right monitor and laptop, prompting them to go back and inspect; (d) receives one stall notification from the left-monitor dashboard and inspects the flagged session on the right monitor where two sessions sit side by side, while an idle task from a previous wave remains untouched on the laptop; and (e) tiles five sessions on the right monitor, with the ticket open in VSCode on the laptop and a dependent task held back for a later execution wave. Participants and surrounding room details have been masked for privacy and visual clarity.}
    \label{fig:user-workspaces}
    \vspace{-20px}
\end{figure*}

Figure~\ref{fig:user-workspaces} illustrates how five different participants used
\pp's three components within the study's three-screen workspace
(two external monitors and a laptop screen).
With the {planning interface}, participants reviewed task
dependencies and execution plans alongside project tickets and
code~(a). With the {run-logger}, they launched agent sessions
with automatic activity capture while retaining their familiar
VS Code and terminal interfaces~(b). With the
{ambient dashboard}, they kept an overview on the left
external monitor while using the right monitor and laptop for
individual agent sessions and code inspection~(c--e). These
arrangements supported monitoring parallel activity, investigating
a potentially stalled session, and coordinating dependent
follow-up work.

\newpage
\subsection{\textbf{\hypidPaper{[RQ2 $\rightarrow$ RQ1]} Exploratory Associations Between Supervisory
Support and Efficiency Gains}}

\label{app:exploratory}
\begin{figure}[!h]
    \centering
    \includegraphics[width=1\linewidth]{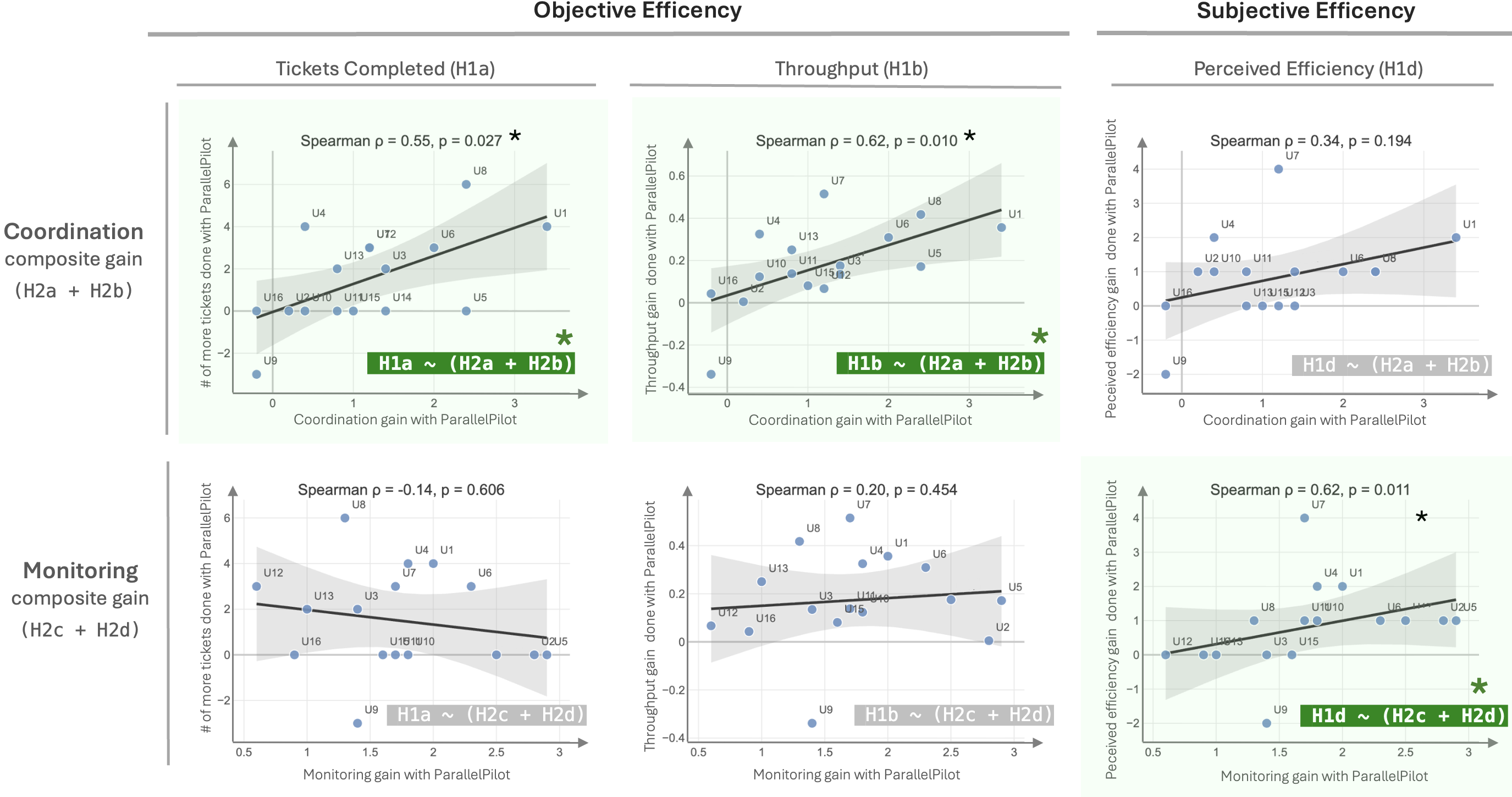}
\caption{Exploratory associations between participant-level gains
in coordination (top row) and monitoring (bottom row) and gains
in ticket completion, throughput, and self-rated efficiency ($N=16$). Each point represents one participant.
Annotations report Spearman correlations and unadjusted
$p$-values; green panels and asterisks indicate $p<.05$. Black lines show descriptive
OLS fits, with gray 95\% confidence bands for the fitted means.}
\Description{Six scatterplots arranged in two rows and three
columns. The top row plots coordination-composite gains against
ticket-completion, throughput, and self-rated-performance gains;
the bottom row plots monitoring-composite gains against the same
outcomes. Blue points are labeled with participant IDs.
Coordination correlations are .55, .62, and .34, respectively;
monitoring correlations are -.14, .20, and .62. Green backgrounds
highlight coordination with ticket completion and throughput,
and monitoring with self-rated performance, based on unadjusted
p-values below .05. Black fitted lines and gray confidence bands
show the direction and uncertainty of descriptive linear trends.
All axes represent condition differences, with positive values
indicating improvement with ParallelPilot.}
    \label{fig:rq2-rq1-correlations}
    \vspace{-20px}
\end{figure}

As shown in Figure~\ref{fig:rq2-rq1-correlations}, we conducted exploratory Spearman correlations to examine whether
participant-level gains in \textit{coordination} and \textit{monitoring} were
associated with gains in ticket completion (\hypidPaper{H1a}), throughput (\hypidPaper{H1b}), and
perceived efficiency (\hypidPaper{H1d}). We constructed
composites by averaging all six coordination items (\hypidPaper{H2a, H2b}) and all eight monitoring
items (\hypidPaper{H2c, H2d}). Gains were calculated as the
with-\pp{} score minus the baseline score. 

Coordination-composite gains were associated with increases in objective efficiency metrics: tickets completed (\hypidPaper{H1a}; $\rho=.55$, $p<0.05$) and tickets completed per minute (\hypidPaper{H1b}; $\rho=.62$, $p<0.05$). Participants' accounts emphasized organizing work to avoid interference and reconciliation: U4 noted the clearer assignment and dependency isolation definitely reduced conflict risk when parallelizing, while U6 shared ``\textit{[\pp's] workspace, worktree isolation helps agents [...] not collide with each other.}'' These accounts suggest a possible link to objective productivity: avoiding conflicting work may reduce the rework needed to complete tickets.

Monitoring-composite gains, in comparison, were associated with perceived efficiency gains (\hypidPaper{H1d}; $\rho=.62$, $p<0.05$), but no significant associations were detected with objective productivity gains. Participants emphasized spending less effort checking progress: U4 described being able to ``\textit{keep track of progress and manage multiple contexts in one place},'' while U1 explained, ``\textit{[in my original workflow], I have to open many windows to track what finished, but since the dashboard already tracks it, I don't think I need to.}'' 
U10 described seeing all three agents finish as ``\textit{the most satisfying UX feature},'' because it made their status immediately visible and allowed them to move on without unnecessary waiting. Together, these accounts suggest that monitoring support primarily improved the experience of supervising parallel work by reducing tracking effort and uncertainty about agent status, rather than directly increasing task output.

\section{Study Instruments}
\subsection{Formative Study -- Survey Questions}
\label{app:formative-survey-questions}

\begin{enumerate}
    \setcounter{enumi}{0}
    \item \textbf{Participant ID.} Please enter the participant ID provided by the researcher.
    \item \textbf{Age} (optional). What is your age? \\
    18--24, 25--34, 35--44, 45--54, 55--64, 65+
    \item \textbf{Gender} (optional). What is your gender? \\
    Woman, Man, Non-binary, Prefer not to say, Other
    \item \textbf{Current Role.} What job title(s) best describe you currently? (You may list multiple roles, e.g., ``Research Intern and PhD Student'' if that applies.)
    \item \textbf{Coding Experience.} How many years of coding experience do you have? \\
    Less than 1 year, 1--3 years, 4--6 years, 7--10 years, More than 10 years, Other
     \item \textbf{AI Coding Tools Used.} Which AI coding tools do you regularly use? (Select all that apply.) \\
    CLI--Claude Code, CLI--Codex, CLI--Copilot, IDE--Claude, IDE--Copilot, IDE--Cursor, Web/Desktop--Codex, Web/Desktop--ChatGPT, Web/Desktop--Claude, Web/Desktop--Copilot, Other
    \item \textbf{Frequency of AI Coding Assistant Usage (General).} How often do you use AI coding assistants in your coding workflow, including using just one agent at a time? \\
    More than 3 times a day, 1--3 times per day, 1--3 times per week, 1--3 times per month, Fewer than once a month
    \item \textbf{Frequency of Parallel AI Coding.} How often do you run multiple AI coding sessions in parallel? \\
    More than 3 times a day, 1--3 times per day, 1--3 times per week, 1--3 times per month, Fewer than once a month
    \item \textbf{Experience with Parallel AI Coding.} Approximately how many times have you used multiple AI coding sessions in parallel? \\
    1--10 times, 10--30 times, More than 30 times
    
    \item \textbf{Typical Number of Concurrent Sessions.} When working in parallel, how many AI coding sessions are typically active at once? (Please provide a range or a number.)
    \item \textbf{Maximum Number of Concurrent Sessions.} What is the largest number of AI coding sessions you have had active at the same time? (Please provide a number.)
    \item \textbf{Current Coding Workflow Distribution.} Thinking about your current coding practice, approximately what percentage of your work falls into each of the following categories? (Example: 40\%, 40\%, 20\%. Then you can simply enter: 40, 40, 20. Please ensure your responses add up to 100\%.   )
    \begin{itemize}
        \item Coding without any AI tools: \_\_\%
        \item Coding with a single AI coding tool/session at a time: \_\_\%
        \item Coding with multiple AI coding sessions in parallel: \_\_\%
    \end{itemize}
  
    \item \textbf{Willingness to Share a Real Multi-Session Project.} During the interview, we will ask you to (screen)share and walk us through a real project where you used multiple AI coding sessions in parallel. To help prepare, please think of a project that you can show and discuss. During the interview, we will ask about:
    \begin{itemize}
        \item The background, context, and goals of the project.
        \item How many AI sessions you used and how work was divided across sessions.
        \item How you monitored, evaluated, and intervened in different sessions over time.
        \item The benefits, challenges, surprises, or concerns you experienced while managing multiple sessions.
    \end{itemize}
    Anything you choose to share will only be viewed by the research team. You may skip any project, repository, file, or detail you are not comfortable showing. Screen sharing will be used only to help the research team understand your workflow; we will not share your screen, code, or projects outside the research team. Any identifying or sensitive information (e.g., repository names, file names, proprietary code, or internal information) will be removed, anonymized, or blurred before appearing in notes or write-ups. You may stop screen sharing or skip any content at any time.

    \emph{Do you have a project in mind that you would be willing to walk us through during the interview?} 
    \begin{itemize}
        \item  Yes, I have a project in mind and would be willing to walk through it during the interview. 
        \end{itemize}

\end{enumerate}
\subsection{Formative Study --  Interview Protocols}
\label{app:formative-interview-protocol}

For this study, an \emph{AI coding session} was defined for participants as one ongoing interaction with an AI coding tool that has its own task context, for example a single Claude Code terminal, a Copilot or Cursor IDE chat, or one AI coding interface working on a specific task. We were especially interested in moments where more than one such session was active at the same time, even if the sessions were at different stages or required different levels of attention.

\subsubsection{Warm-Up and Background}
\begin{itemize}
    \item Tell me a bit about your role and the kind of coding work you do day to day.
    \item When did you start using AI coding tools? What has your experience been like so far, and which tools or setup have you found most useful?
    \item What has your experience been like using multiple AI coding sessions at the same time?
\end{itemize}

\subsubsection{Why Parallel}
\begin{itemize}
    \item \textit{Pros:} Why do you run multiple AI coding sessions at the same time instead of one after another?
    \item \textit{Cons:} What are the downsides or coordination costs of using multiple sessions at once?
    \item \textit{Tradeoff:} When is using multiple sessions worth the effort, and when would you rather use just one session or none?
    
\end{itemize}

\subsubsection{Task Walkthrough}

Walk me through a recent, real setup where you had several sessions running, not a tidied-up or typical one. 

\subsubsection{Session Roster}
\begin{itemize}
\item What repo was this, and what were you trying to solve? 
    \item \textit{Session roster:} Can you give me a quick map of the sessions that were running? How many were there, and what was each one roughly working on?
    \item \textit{Number rationale:} Why did you need that number of sessions?
    \item \textit{Work division:} How did you decide how to divide the work across sessions?
\end{itemize}

\subsubsection{Session Metadata}
\begin{itemize}
    \item \textit{Session purpose:} What was this session trying to do, and why did you make it a separate session rather than keeping it inside another session?
    \item \textit{Work type:} Which best describes the work this session was doing?
\end{itemize}

\subsubsection{Attention and Awareness}
\begin{itemize}
    \item \textit{When to check \& reading strategy:} What do you actively monitor, and what do you usually skim or ignore? Did you read what the session was doing as it worked, or mostly inspect the result afterward?
    \item \textit{Warning signals:} What kinds of things trigger you to step in, for example errors, weird output, drift, conflicts, or just a gut feeling?
    \item \textit{Legibility of status:} Was anything important hard to see? Did this session have a clear signal for ``everything is fine'' versus ``you need to inspect this carefully''?
    \item \textit{Evaluation ability:} Did you know what a good result should look like, and how to evaluate whether the session was doing well? Were there parts where you felt unable to judge the quality of the work?
\end{itemize}

\subsubsection{Closing Questions}
\begin{itemize}
    \item Tell me about the last time you intervened in an agent's work.
    \item What's the maximum number of concurrent sessions that feels useful, and what limits that number?
    \item If these agents were people, what management practices from managing humans carry over?
\end{itemize}

\subsection{\pp User Study -- Pre-Survey Questions}
\label{app:pre-survey}

The pre-survey collected participants' backgrounds, prior experience
with parallel AI coding, and preferred screen configurations.
Consent questions and name-collection fields are omitted below.
Question numbers follow the original survey. All questions were
required unless marked optional.

\begin{enumerate}[start=3, leftmargin=*, itemsep=0.75em]
    \item \textbf{Participant ID.}
    Please enter the participant ID provided by the researcher.

    \textit{Response:} Free text.

    \item \textbf{Age (optional).}
    What is your age?

    \textit{Options:} 18--24; 25--34; 35--44; 45--54; 55--64; 65+.

    \item \textbf{Gender (optional).}
    What is your gender?

    \textit{Options:} Woman; Man; Non-binary; Prefer not to say; Other.

    \item \textbf{Coding experience.}
    How many years of coding experience do you have?

    \textit{Options:} Less than 1 year; 1--3 years; 4--6 years;
    7--10 years; More than 10 years; Other.

    \item \textbf{Frequency of AI coding assistant usage (general).}
    How often do you use AI coding assistants in your coding workflow
    (including using just one agent at a time)?

    \textit{Options:} More than 3 times a day; 1--3 times per day;
    1--3 times per week; 1--3 times per month;
    Fewer than once a month; Other.


    \textit{Options:} 1--10 times; 10--30 times;
    More than 30 times; Other.

    \item \textbf{Expertise with parallel AI coding.}
    Please select the group that best describes your parallel AI
    coding experiences.

    \begin{itemize}[leftmargin=*, nosep]
        \item \textbf{Beginner:} I have only tried parallel AI coding
        a few times or am just getting started.
        \item \textbf{Somewhat experienced:} I have used parallel AI
        coding sessions occasionally and am still developing my approach.
        \item \textbf{Experienced:} I regularly use parallel AI coding
        sessions and have established my own methods.
        \item \textbf{Expert:} I know how to set up parallel AI coding
        sessions, teach others, and have developed my own methods.
    \end{itemize}

    \item \textbf{Current experiences with parallel AI coding.}
    Reflecting on your current use of parallel AI coding sessions,
    please indicate how much you agree or disagree with each
    statement below.

    \textit{Scale:} 1 = Strongly disagree;
    4 = Neither agree nor disagree; 7 = Strongly agree.

    \begin{enumerate}[label=(\alph*), leftmargin=*, nosep]
        \item I find it difficult to break down coding goals into
        tasks for separate AI sessions.
        \item I struggle to determine which tasks can be run in
        parallel versus those that require sequential execution.
        \item I often realize too late that two agents worked on
        overlapping or conflicting code.
        \item It is challenging to keep track of what each agent
        is doing simultaneously.
        \item I often miss when an agent finishes, stalls,
        or becomes blocked.
        \item Switching between sessions or tabs to check status
        feels mentally taxing.
        \item I spend time figuring out which agent needs my
        attention next.
        \item I rely on reviewing raw terminals or logs to
        understand agent activity.
        \item After returning to a session, it is hard to recall
        what was previously accomplished.
        \item I find it difficult to reconstruct why an agent
        made certain decisions.
        \item I feel less in control when multiple agents are
        running at once.
        \item I am unsure when to intervene, redirect, or stop
        an agent.
    \end{enumerate}

    \item \textbf{Preferred working setup.}
    What screen setup do you actively use for (coding and)
    parallel AI coding?

    S = Laptop/Mac screen, H = Horizontal monitor,
    V = Vertical monitor. Only count screens actively used for
    coding or coding-related tasks; do not count screens used
    only for non-coding activities (e.g., Teams, Spotify, YouTube).

    \textit{Response:} Multiple selection.

    \textit{Options:} S only; S + H; S + V; S + 2H; S + H + V;
    No preference---I am comfortable with any setup; Other.

    \item \textbf{Additional comments (optional).}

    \textit{Response:} Free text.
\end{enumerate}

\subsection{\pp User Study --  Post-Condition Survey Questions}
\label{app:post-condition-survey}

Participants completed this survey after each coding condition.
The post-condition and comparative surveys were presented in the
same form; their original question numbers are retained here.

\paragraph{Instructions.}
Please rate the following statements based on your experience
during this AI coding session. Use a scale from 1 (Strongly disagree)
to 7 (Strongly agree), unless otherwise specified.
Please share your thoughts aloud while completing this survey.

\begin{enumerate}[leftmargin=*, itemsep=0.75em]
    \item \textbf{Participant ID.}

    \textit{Response:} Free text.

    \item \textbf{Session task.}
    Select the project you worked on during this session.

    \textit{Options:}
    Sparkmatch;
    Kart.

    \item \textbf{Session condition.}
    Which condition did you experience during this session?

    \textit{Options:} Without ParallelPilot; With ParallelPilot.



    \item \textbf{Supervision-specific load.}
    Please rate the following statements about supervising
    agents/sessions.

    \textit{Scale:} 1 = Strongly disagree; 7 = Strongly agree.

    \begin{enumerate}[label=(\alph*), leftmargin=*, nosep]
        \item Keeping track of what every agent was doing took
        a lot of mental effort.
        \item Switching between agents/sessions was disruptive
        to my focus.
        \item I felt like I was constantly context-switching
        rather than making progress.
    \end{enumerate}

    \item \textbf{Awareness, memory, and recall.}

    \textit{Scale:} 1 = Strongly disagree; 7 = Strongly agree.

    \begin{enumerate}[label=(\alph*), leftmargin=*, nosep]
        \item At any moment, I knew what each agent was working on.
        \item I noticed when an agent finished, stalled,
        or needed my input.
        \item I had a clear overall picture of progress across
        all sessions at once.
    \end{enumerate}

    \item \textbf{Planning, decomposition, and delegation.}

    \textit{Scale:} 1 = Strongly disagree; 7 = Strongly agree.

    \begin{enumerate}[label=(\alph*), leftmargin=*, nosep]
        \item I had a clear plan for how to break the work
        into parallel pieces.
        \item I understood the dependencies between the pieces
        of work I assigned.
        \item The way I assigned tasks to agents matched what
        each task actually needed.
        \item Agents ended up working on overlapping or
        conflicting parts of the code.
        \item I felt confident delegating work to agents
        without watching them closely.
    \end{enumerate}

    \item \textbf{Control, verification, and intervention.}

    \textit{Scale:} 1 = Strongly disagree; 7 = Strongly agree.

    \begin{enumerate}[label=(\alph*), leftmargin=*, nosep]
        \item I felt in control of what the agents were doing.
        \item When I intervened, it successfully (re)directed
        the agent.
        \item I felt confident stopping, correcting, or
        steering an agent.
        \item I trusted the output produced in this block.
    \end{enumerate}

    \item \textbf{Usability and experience.}

    \textit{Scale:} 1 = Strongly disagree; 7 = Strongly agree.

    \begin{enumerate}[label=(\alph*), leftmargin=*, nosep]
        \item I found this setup easy to use.
        \item The setup's capabilities met my needs for
        supervising multiple agents.
        \item I needed to do a lot of manual bookkeeping
        to stay on top of things.
        \item I felt efficient working this way.
        \item I would want to work this way again on real tasks.
    \end{enumerate}

    \item \textbf{Comfortable number of supervised agents/sessions.}
    Under this setup, how many agents or sessions do you feel
    you could comfortably supervise at the same time?
    Select all that apply.

    \textit{Options:} 1; 2; 3; 4; 5; 6; 7; 8+.

    \item \textbf{Session ID.}

    \textit{Options:} First Task Done; Second Task Done
\end{enumerate}

\subsection{\pp User Study -- Final Comparative Survey Questions}
\label{app:comparative-survey}

The final portion of the experience survey collected overall
evaluations of ParallelPilot, feature usefulness ratings,
and comparisons between the two conditions.
Question 12 was required; Questions 13--16 were not marked required.

\begin{enumerate}[start=11, leftmargin=*, itemsep=0.75em]
    \item \textbf{Overall evaluation of ParallelPilot.}
    Please rate your agreement with each statement about your
    experience using ParallelPilot during this task.

    \textit{Scale:} 1 = Strongly disagree / Very negative;
    4 = Neither agree nor disagree / Neutral;
    7 = Strongly agree / Very positive.
    Item-specific anchors are given below where applicable.

    \begin{enumerate}[label=(\alph*), leftmargin=*, nosep]
        \item Overall, how would you rate your experience
        using ParallelPilot during this task?
        (1 = Very negative, 7 = Very positive)
        \item Overall, how satisfied were you with ParallelPilot?
        (1 = Very dissatisfied, 7 = Very satisfied)
        \item ParallelPilot helped me manage multiple AI coding
        sessions more effectively.
        \item I would use a tool like ParallelPilot for future
        parallel AI coding tasks.
        \item ParallelPilot improved my confidence in supervising
        multiple AI agents simultaneously.
    \end{enumerate}

    \item \textbf{ParallelPilot usefulness.}
    Please rate the usefulness of the tool features.

    \textit{Scale:} 1 = Not at all useful; 4 = Neutral;
    7 = Very useful; additional option: Didn't use.
    The survey instructions described the upper endpoint
    as ``Extremely useful,'' while the response grid labeled
    it ``Very useful.''

    \begin{enumerate}[label=(\alph*), leftmargin=*, nosep]
        \item Plan / task decomposition view.
        \item Workspace or branch isolation per agent.
        \item Session logs / event history.
        \item Live monitoring dashboard / status view.
    \end{enumerate}

    \item \textbf{Setup/condition preference.}
    Please select one option for each aspect.

    \textit{Options:} Strongly prefer no-tool;
    Slightly prefer no-tool; No preference;
    Slightly prefer ParallelPilot; Strongly prefer ParallelPilot.

    \begin{enumerate}[label=(\alph*), leftmargin=*, nosep]
        \item Overall preference.
        \item Which setup made you feel more in control?
        \item Which setup made it easier to stay aware of all agents?
        \item Which setup made it easier to plan and split the work?
        \item Which setup made it easier to verify the output?
        \item Which setup felt less mentally taxing?
        \item If we assume both tasks were equally hard, in which
        block do you think you produced better-quality work?
        \item If we assume both tasks were equally hard, in which
        block do you think you produced more work?
    \end{enumerate}

    \item \textbf{Task difficulty.}
    Was one task noticeably harder than the other?

    \textit{Options:} About equal;
    Yes, Kart is harder than Sparkmatch;
    Yes, Sparkmatch is harder than Kart.

    \item \textbf{Additional comments.}
    Anything else you'd like us to know?

    \textit{Response:} Free text.
\end{enumerate}

\subsection{\pp User Study --  Interview Protocols}
\label{app:interview}

\subsubsection{Post-task interview}

The research protocol included the following open-ended
post-task interview questions.

\begin{enumerate}[leftmargin=*, itemsep=0.2em]
    \item Compared with your normal workflow, what did the
    prototype make easier or harder?

    \item Did the prototype help you remember what each session
    was doing? Give a concrete example.

    \item Did the prototype help you notice risks, dependencies,
    conflicts, or places that needed verification?

    \item Did the prototype change when you chose to intervene
    or let a session continue?

    \item Which condition felt faster, safer, more controlled,
    or more mentally demanding?

    \item Would you use something like this in your own workflow?
    What would need to change first?
\end{enumerate}

\subsubsection{Think-aloud prompts during the tasks}

The protocol also included the following prompts for use
during the coding sessions.

\begin{enumerate}[leftmargin=*, itemsep=0.2em]
    \item What are you trying to get this AI session to do right now?

    \item What information are you using to decide whether
    this session is on track?

    \item Which session needs your attention next, and why?

    \item What would make you intervene, redirect,
    or stop this session?

    \item If you came back after a few minutes away,
    what would you look at first to recover context?
\end{enumerate}

\end{document}